\documentclass[twocolumn,times]{aastex62}

\usepackage{amsmath,amstext}

\usepackage{newtxmath} 

\usepackage{epsfig}

\usepackage{float}
\usepackage{natbib}
\usepackage[figure,figure*]{hypcap}

\newcommand\chandra{\emph{Chandra}}
\newcommand\msun{$\mathrm{M}_{\odot}$}

\defcitealias{2000ApJ...541..542M}{M00}
\defcitealias{2007PhR...443....1M}{MV07}
\defcitealias{2006ApJ...650..102A}{A06}
\defcitealias{2013MNRAS.436.1721R}{R13}

\shorttitle{Cold Fronts of A2142}
\shortauthors{Wang \& Markevitch}

\begin{document}

\title{A DEEP X-RAY LOOK AT ABELL 2142 --- VISCOSITY CONSTRAINTS FROM
KELVIN-HELMHOLTZ EDDIES,\\
A DISPLACED COOL PEAK THAT MAKES A WARM CORE, AND A
POSSIBLE PLASMA DEPLETION LAYER}

\author{Qian H. S. Wang}
\affiliation{Department of Astronomy, University of Maryland, College Park, MD
20742}

\author{Maxim Markevitch}
\affiliation{Astrophysics Science Division, NASA Goddard Space Flight Center,
Greenbelt, MD 20771}
\affiliation{Joint Space-Science Institute, University of Maryland, College
Park, MD, 20742}

\begin{abstract}

We analyzed 200 ks of \chandra\ ACIS observations of the merging galaxy
cluster A2142 to examine its prominent cold fronts in detail. We find that the
southern cold front exhibits well-developed Kelvin-Helmholtz (KH) eddies seen
in the sky plane. Comparing their wavelength and amplitude with those in
hydrodynamic simulations of cold fronts in viscous gas, and estimating the gas tangential
velocity from centripetal acceleration, we constrain the effective viscosity
to be at most 1/5 of Spitzer isotropic viscosity, but consistent with full Braginskii
anisotropic viscosity for magnetized plasma. While the northwestern front does
not show obvious eddies, its shape and the structure of its brightness profile
suggest KH eddies seen in projection. The southern cold front continues in a
spiral to the center of the cluster, ending with another cold front only 12
kpc from the gas density peak. The cool peak itself is displaced $\sim$30~kpc
from the BCG (the biggest such offset among centrally-peaked clusters), while
the X-ray emission on a larger scale is still centered on the BCG, indicating
that the BCG is at the center of the gravitational potential and the cool gas
is sloshing in it. The specific entropy index of the gas in the peak
($K\approx49$~keV~cm$^2$) makes A2142 a rare ``warm core''; apparently the
large displacement of the cool peak by sloshing is the reason.  Finally, we
find a subtle narrow, straight channel with a 10\% drop in X-ray brightness,
aligned with the southern cold front --- possibly a plasma depletion layer in
projection.\\

\end{abstract}

\section{Introduction}

The phenomenon known as a ``cold front'' was first discovered by \chandra\ in
the galaxy clusters A2142 \citep{2000ApJ...541..542M} (hereafter M00) and
A3667 \citep{2001ApJ...551..160V,2001ApJ...549L..47V}. Cold fronts are contact
discontinuities in the density and temperature of the intracluster gas, seen
in the sky plane as sharp edges (discontinuities of the gradient) of the X-ray
brightness, usually unresolved even with the \chandra\ angular resolution (for
a review see \citealt{2007PhR...443....1M}, hereafter MV07).  Cold fronts may
look similar to shocks in cluster X-ray images, but the gas temperature jump
has the opposite sign --- in the cold front, the temperature is lower on the
denser side, so the two sides are near (though not exactly in) pressure
equilibrium. Unlike in shock fronts, there is no flow of gas across the cold
front, but there is often a tangential velocity difference.

Cold fronts can form during a merger as a result of ram pressure stripping of
the infalling subcluster (the original proposal for A2142 in
\citetalias{2000ApJ...541..542M}). Clear examples of such fronts are the
Bullet subcluster \citep{2002ApJ...567L..27M} and the infalling galaxy
NGC\,1404 (\citealt{2005ApJ...621..663M}; \citealt{2017ApJ...834...74S}).
Another class of cold fronts is observed in or near most cool cores, often as
multiple concentric edges in a spiral pattern. These edges are caused by an
off-axis subcluster merger and the resulting displacement of the dense core
gas from the minimum of the gravitational potential, which sets off
long-lasting sloshing of that gas in the potential well
(\citealt{2001ApJ...562L.153M}; \citealt{2006ApJ...650..102A}, hereafter A06;
\citetalias{2007PhR...443....1M}). Such fronts are found in most cool cores
\citep{2010A&A...516A..32G}, even in otherwise relaxed clusters; examples are
RXJ1720.1+26 \citep{2001ApJ...555..205M}, A2029 \citep{2004ApJ...616..178C},
Ophiuchus (\citetalias{2006ApJ...650..102A}; \citealt{2010MNRAS.405.1624M};
\citealt{2010ApJ...717..908Z}; \citealt{2016MNRAS.460.2752W}; A496
\citep{2007ApJ...671..181D}, Perseus (\citealt{2003ApJ...590..225C};
\citealt{2012ApJ...757..182S}), Virgo (\citealt{2010MNRAS.405...91S};
\citealt{2011MNRAS.413.2057R}; \citealt{2016MNRAS.455..846W}), and, as we now
believe (\citealt{2005ApJ...618..227T}; \citetalias{2007PhR...443....1M}),
A2142.

Both types of cold fronts can be used for interesting tests of the
microphysics of the intracluster plasma (\citetalias{2007PhR...443....1M}). In
particular, the abruptness of the temperature and density changes across the
front strongly limits thermal conductivity and diffusion
(\citealt{2000MNRAS.317L..57E}; \citealt{2001ApJ...551..160V};
\citetalias{2007PhR...443....1M}), suggesting that magnetic field drapes
around the front surface and insulate the front. Because the gas tangential
velocity is discontinuous across the front, cold fronts should develop
Kelvin-Helmholtz (KH) instabilities. As indeed observed in, e.g., A3667,
Bullet and NGC\,1404, they lead to eventual dissolution of the sharp
interface. The growth of KH instability depends on --- and therefore can be
used to constrain --- the plasma viscosity and the structure and strength of
the magnetic fields (\citealt{2001ApJ...549L..47V};
\citetalias{2007PhR...443....1M}; \citealt{2013MNRAS.436.1721R}, hereafter
R13), though separating these two stabilizing effects may not be
straightforward \citep{2015ApJ...798...90Z}. Evidence for cold fronts
developing KH instabilites has been seen indirectly in the form of multi-edge
structure of the radial brightness profile and ``boxy'' shape of the fronts,
both consistent with being KH eddies seen in projection (e.g., Virgo, A496,
\citealt{2013ApJ...764...60R}, \citealt{2012MNRAS.420.3632R}; NGC\,1404,
\citealt{2017ApJ...834...74S}; A3667, \citealt{2017MNRAS.467.3662I}). Their
existence has been used to place an upper limit on the plasma {\em
isotropic}\/ viscosity (that is, disregarding the effect of the magnetic
fields) to be $\sim$10\% of the Spitzer value. As shown by MHD simulations
\citep{2015ApJ...798...90Z}, in the context of sloshing cold fronts, the
suppression of KH instabilities in a plasma with a magnetic field draping
around the cold front, with anisotropic Braginskii viscosity, should be
qualitatively similar to the effect of a 1/10 Spitzer isotropic viscosity.

So far, KH eddies in the plane of the sky have been seen only in A3667
(\citealt{2002ApJ...569L..31M}; \citealt{2011scgg.conf...53V};
\citealt{2017MNRAS.467.3662I}). A possible eddy has also been reported at a sloshing cold front in Perseus \citep{2017MNRAS.468.2506W}, although the Perseus core is full of AGN bubbles and that feature could also be one of those. Those are the ones that can provide the most
unambiguous constraints on the plasma microphysics. In this paper, we present
another example of a cold front that shows apparent KH eddies, the southern
front in A2142, based on a deeper \chandra\ observation of the cluster core.
In addition, we analyze a recently found cold front at a very small radius, as
well as two other interesting effects: a cool peak displaced from the central
galaxy, as well as a subtle channel in the cluster X-ray brightness --- a
phenomenon similar to that we have recently discovered in another cluster,
A520 \citep{2016ApJ...833...99W}.

While we concentrate on the core of A2142, where we now observe 3 concentric
cold fronts (at $r\approx 12-340$~kpc), this cluster exhibits another cold
front far outside the core, 1~Mpc from the center \citep{2013A&A...556A..44R},
outside the \chandra\ coverage. A set of multiple concentric fronts at such
different radii indicates ``an extreme case of sloshing'', quoting the above
authors. Interestingly, A2142 has a specific entropy in the gas density peak
that makes it a relatively rare ``warm core'' --- intermediate between
cool-core and non-cool-core clusters (\citealt{2009ApJS..182...12C};
\citealt{2017ApJ...841...71G}). We will try to clarify if this can be related
to the observed strong sloshing. A2142 also has a giant radio halo whose
structure spatially correlates with the cold fronts on all scales
\citep{2017A&A...603A.125V}.

In \autoref{sec:xrayanalysis}, we describe our treatment of \chandra\ data, as
well as spectral and imaging analyses. In \autoref{sec:temperaturemap}, we
describe the procedure we used to generate a wavelet enhanced temperature map
of the cluster's central regions. In \autoref{sec:thecoldfronts}, we describe
each of the three cold fronts in turn, including the displacement of the cool
core from the BCG in \autoref{sec:showinner}. We then discuss in
\autoref{sec:kh} our results in the context of constraining viscosity, and in
\autoref{sec:channel} a possible plasma depletion sheet. Finally we summarize
our results in \autoref{sec:summary}.

At the cluster redshift of $z=0.089$, 1\arcsec\ is 1.66~kpc for $h=0.7$ and
$\Omega_M = 0.3$. Unless otherwise stated, errors in the text are given at
90\% confidence.

\section{X-ray Data Analysis}
\label{sec:xrayanalysis}

We combined the archival \chandra\ Advanced CCD Imaging Spectrometer (ACIS)
observations with ObsID 5005, 15186, 16564, and 16565, omitting for
convenience the short (16~ks) dataset analyzed in
\citetalias{2000ApJ...541..542M}. ObsID 5005 (45~ks) was taken in 2005 (PI L.
VanSpeybroeck) and had the cluster center in ACIS-I3; it has been analyzed by
\cite{2009ApJ...704.1349O} and \cite{2011PhDT........14J}. The latter three
(153~ks total) were taken in 2014 (PI M. Markevitch) and centered the cluster
in ACIS-S3. An image from these observations have been looked at by
\cite{2016MNRAS.461..684W}. We processed the data using CIAO (v4.9.1) and
CALDB (v4.7.7), with standard event filtering procedure to mask bad pixels,
filter by event grades, remove cosmic ray afterglows and streak events, and
detector background events identified using the VFAINT mode data. The data
were then checked for background flares using the 2.5--7~keV light curve in
1~ks time bins in a cluster-free region, separately for the FI and BI chips.
As a more sensitive check for faint flares, we also used the ratio of
2.5--7~keV to 9.5--12~keV counts. There were no period with strong flares.
The final data we used have a total exposure of 197~ks, which is 97\% of the
raw exposure.

We accounted for the background following \cite{2003ApJ...583...70M} and
\cite{2006ApJ...645...95H}, using the blank-sky data sets from CALDB. For
ObsID 5005, we used the Period E dataset with an exposure of 1.55~Ms. For
ObsIDs 15186, 16564, and 16565, we used the Period F dataset with an exposure
of 800~ks. For both imaging and spectral analysis, the background was scaled
by the ratio of the 9.5--12~keV counts (separately for front-illuminated and
back-illuminated chips), which corrects for the secular background rate
variability. The 90\% uncertainty of the 0.8--9~keV quiescent background
modeled in such a way is 3\% \citep{2006ApJ...645...95H}, so we vary the
background by this amount and include the effect in quadrature in our
temperature measurement errors. The ACIS readout artifact was modeled using
\texttt{make\_readout\_bg}\footnote
{http://cxc.harvard.edu/contrib/maxim/make\_readout\_bg}
and treated as an additional background component, as in
\citetalias{2000ApJ...541..542M}. We identified point sources for exclusion from our analysis
by visual inspection using the 0.8--4~keV and 2--7~keV images at different
binning and smoothing scales.

Spectral analysis was performed in XSPEC (version 12.9.1p). Instrument
responses for spectral analysis were generated as described in
\cite{2005ApJ...628..655V}. We used the CHAV tools to generate the PHA, ARF,
and RMF files for each pointing and then combined the data products. PHA files
from different pointings were coadded for each of the observed data, blank-sky background, and simulated readout background, while ARFs and RMFs were weighed by the counts in the 0.5--2~keV band
(where most of the events are) in the spectral extraction region.

A single-temperature fit to the whole cluster in a 4\arcmin\ circle (0.4~Mpc,
covers most of the S3 chip) centered on $(\alpha, \delta)=$(15:58:20.4,
+27:13:52.7) (FK5, J2000), using the 0.8--9~keV band and the
\texttt{apec*wabs} model, gives $T=8.0\pm0.1$~keV, metal abundance
$0.28\pm0.01$ (relative to \citealt{1989GeCoA..53..197A}), and absorption
column $N_H=(7.7\pm0.3)\times10^{20}$~cm$^{-2}$. The errors are formal errors
from fitting with and the effect of the 3\% uncertainty in the blank-sky
background added in quadrature. The best-fit temperature and abundance are the
same as those reported in \citetalias{2000ApJ...541..542M}, while the best-fit
$N_H$ is double the value $3.8\times10^{20}$~cm$^{-2}$ from the LAB survey
\citep{2005A&A...440..775K}, which probably reflects the uncertainty of the
ACIS calibration at the lowest energies --- the reason for our excluding
$E<0.8$ keV from the fits.
In the analysis below, we fix the abundance and $N_H$\/ to the cluster-wide
best-fit values.

To make the exposure-corrected images, we created exposure maps using Alexey
Vikhlinin's tools%
\footnote{http://hea-www.harvard.edu/~alexey/CHAV},
assuming the spectrum of a single-temperature plasma with best-fit parameters
from the 4\arcmin-radius circle described above.
These are images of effective exposure time that include vignetting and
variations in detector efficiency. Varying the assumed temperature within
the range found in the cluster would make little difference to the broad-band
exposure map, as the counts are dominated by those around the peak of the ACIS
effective area at 1--2~keV. (For narrow-band exposure maps used in
\autoref{sec:temperaturemap} it matters even less.)
We divided the coadded (in sky coordinates) background-subtracted count images
by the coadded exposure maps to get the final flux images.

\begin{figure*}[!ht]
        \centering
        \leavevmode
        \includegraphics[width=170mm]{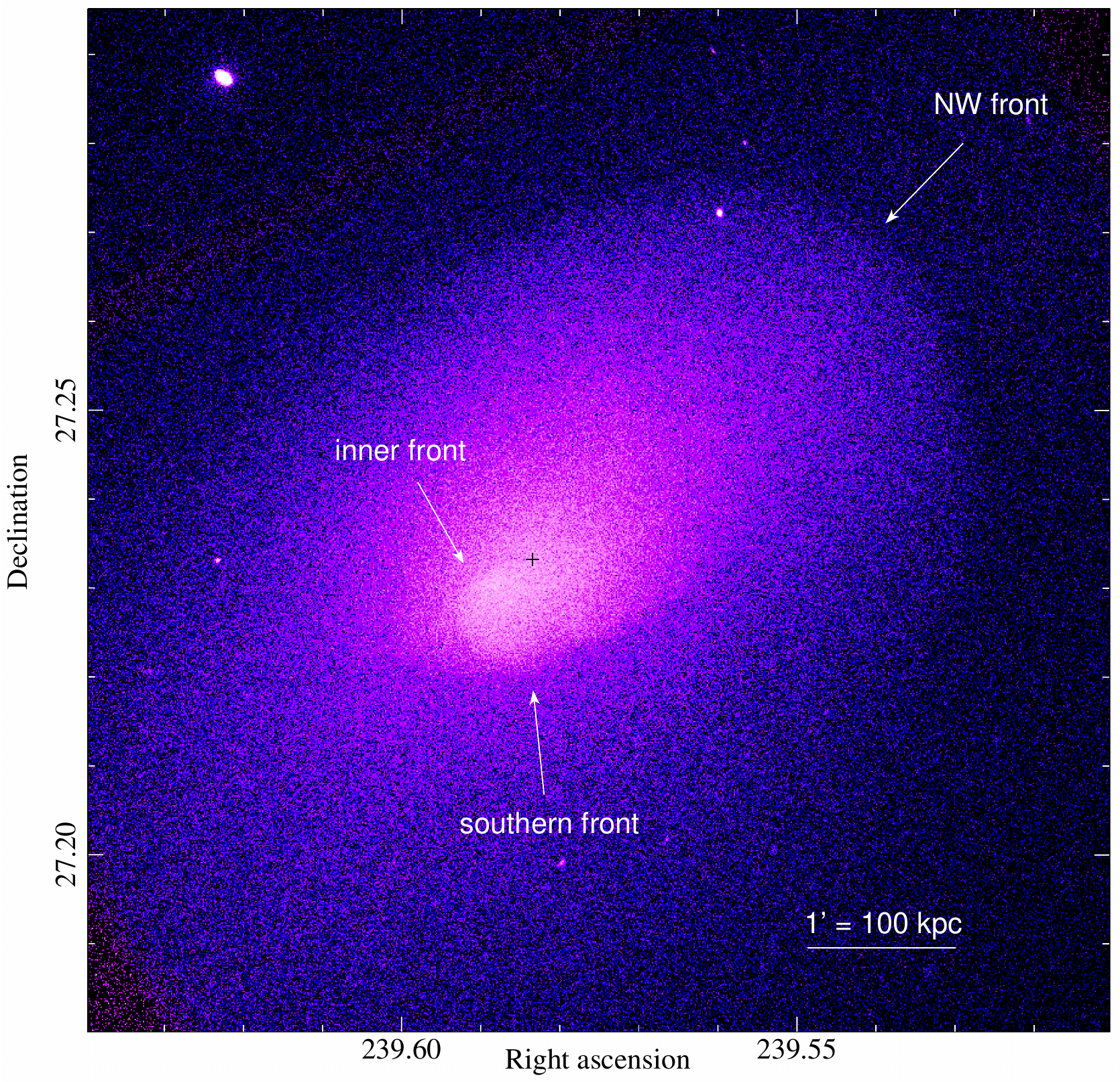}
        \hfil

        \caption{
          A broad view of the features we studied in A2142, shown by an
          unbinned 0.8--4~keV \chandra\ image (1~pixel is 0.5\arcsec). The
          cross marks the position of the BCG.}
        \label{fig:splash}
\end{figure*}

\subsection{Temperature Map of the Small-Scale Structure}
\label{sec:temperaturemap}

To determine the nature of the X-ray structure in the cluster core, we derived
a temperature map of the core by subtracting the smoother, large-scale
emission component, in order to enhances the contrast of the small-scale
features --- that is, to get closer to their true temperatures. Because the
precise 3D geometry of the gas in this asymmetric cluster is unknown, such a
map necessarily provides only a qualitative picture of the core of A2142.

The map shown in \autoref{fig:southern_inner}(a) was derived following the
method described in \citetalias{2000ApJ...541..542M} (without the deprojection
step) and \cite{2016ApJ...833...99W}. We extracted six narrow-band images in
the 0.8--1--1.5--2--4--6--9~keV bands. The flux and error images were smoothed
by wavelets prior to deriving the temperature map, using the same wavelet
decomposition coefficients for all bands.
A single-temperature thermal model was fitted for each pixel to the six flux
values from the narrow band images, fixing the absorption column and metal
abundance to the cluster best-fit values. This resulted in a wavelet-smoothed
temperature map.

Wavelet decomposition separates the structures in the image at different
scales, and helps us qualitatively deproject the large-scale components.
Unlike a smoothing scheme such as Gaussian smoothing, which blurs everything
with a symmetric kernel, wavelet decomposition preserves the shapes and
brightness contrast of interesting small-scale features while at the same time
having a basis in the statistical significance of the structures selected by
the algorithm.  Using a method described in \cite{1994ApJ...435..162V,
1998ApJ...502..558V}, we extracted wavelet components (with the {\it atrous}\/
kernel and scales increasing in geometric progression) from images binned to
1.5\arcsec\ pixels, on scales of 2.5, 5, 10, 20, 39, and 78~kpc (or 1.5, 3,
6, 12, 24, and 47\arcsec). Point sources were remove from the images at
each scale and the components were coadded. Error images were treated with the
same procedure.

\section{Cold Fronts}
\label{sec:thecoldfronts}

\begin{figure*}[!ht]
        \centering
        \leavevmode
        \gridline{
        \fig{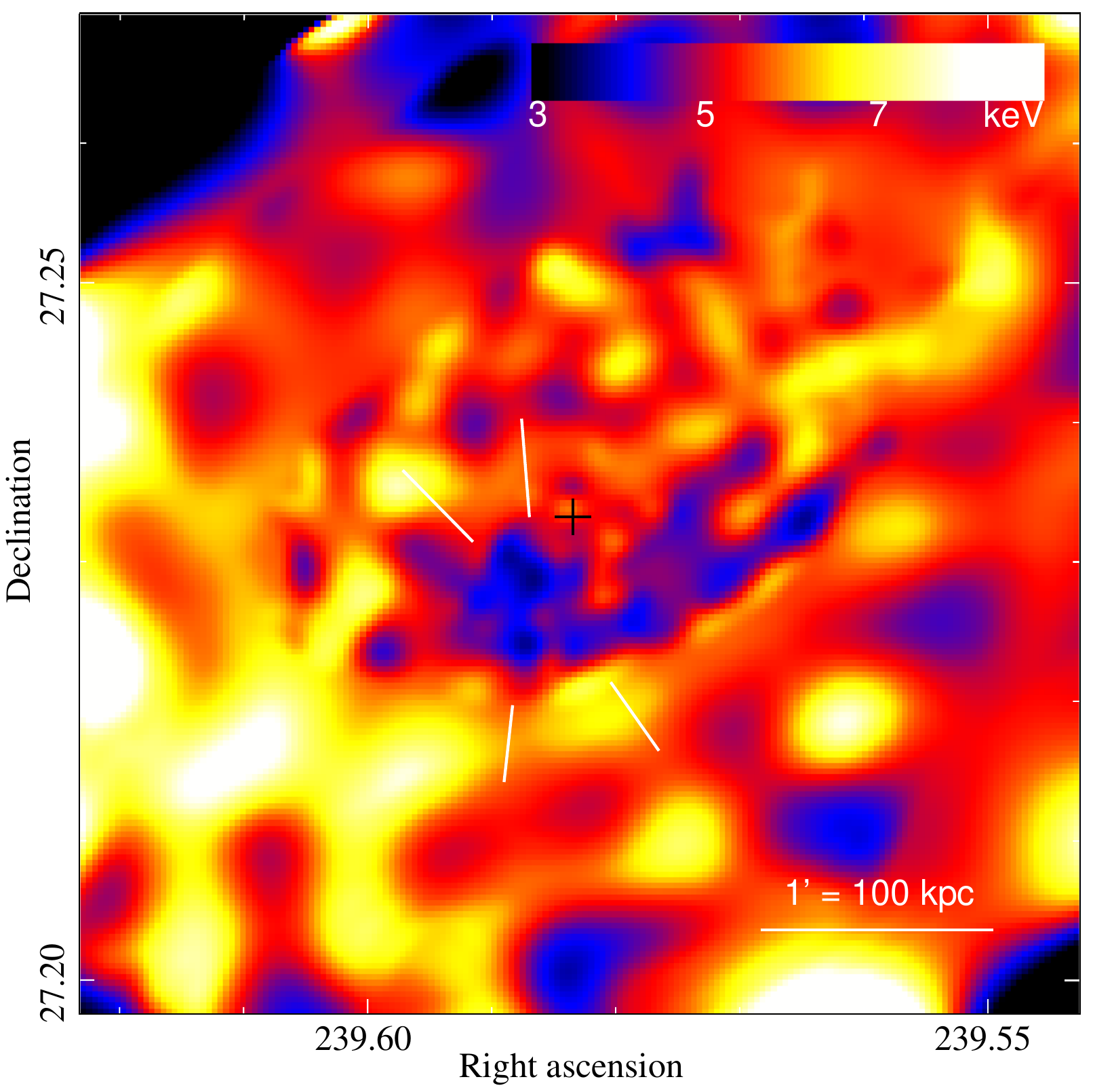}{0.5\textwidth}{(a)}
        \fig{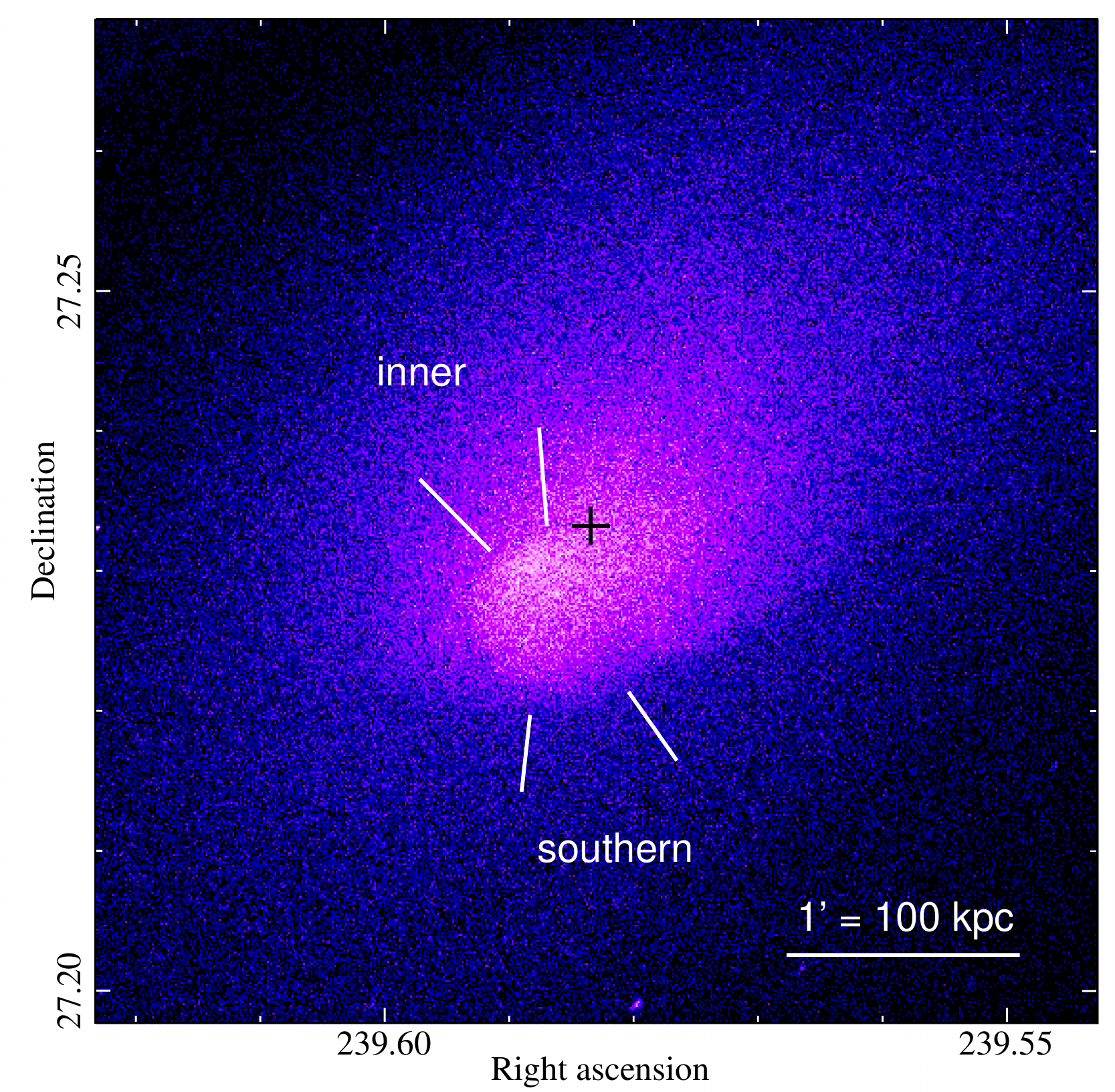}{0.5\textwidth}{(b)}
        }
        \gridline{
        \fig{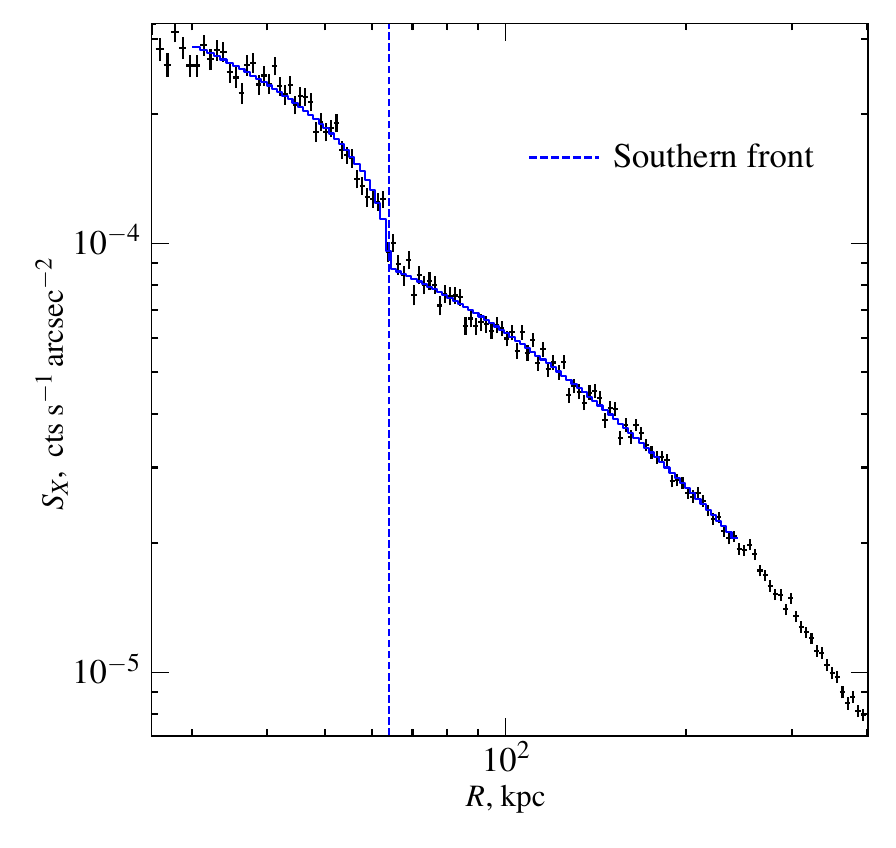}{0.5\textwidth}{(c)}
        \fig{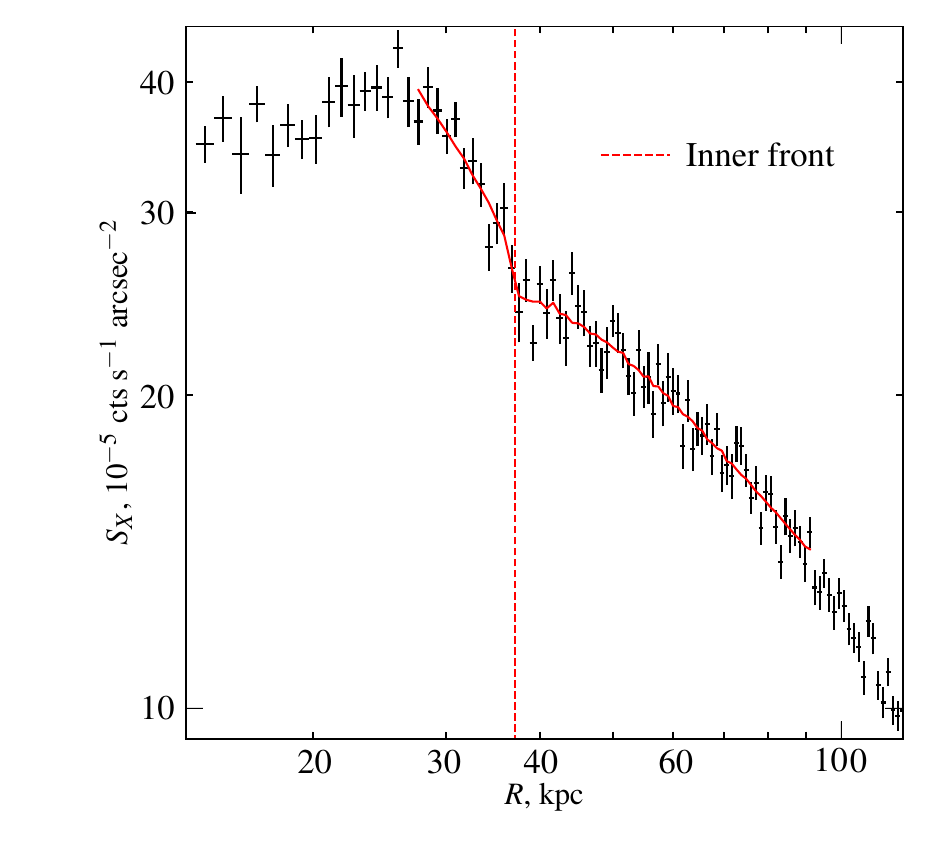}{0.5\textwidth}{(d)}
        }
        \caption{
          (a) Temperature map created using wavelet reconstructed narrow-band
          images, keeping only components on scales up to 47\arcsec\ (=78~kpc).
          This has the effect of deprojecting the larger-scale components for
          a better qualitative view of the temperature structure. A 1\arcsec\
          Gaussian was used to smooth edge artifacts without changing its
          appearance qualitatively.
          (b) 0.8--4~keV image of the same zoom as (a). The white
          lines indicate the width of the sectors used to model the surface
          brightness profiles of the southern and inner cold fronts. The
          cross marks the BCG position.
          (c) X-ray surface brightness profile taken across one of the
          suspected KH eddies in the southern front, in the region shown in
          (b). Blue solid line is the projection of the 3D density model,
          using a power law inside the cold front and a beta model profile
          outside (see \autoref{sec:show-southern} for details). It is drawn for the range of $R$ used in the fitting. The
          dashed line marks the best fit position of the edge.
          (d) Surface brightness profile of the inner front in the sector
          shown in (b). Red solid line is the projection of the 3D density
          model (see \autoref{sec:showinner} for details). The dashed line
          marks the best fit position of the edge.
          }
\label{fig:southern_inner}
\end{figure*}

The 0.8--4 keV A2142 image, full-resolution without any smoothing or
enhancements, is shown in \autoref{fig:splash}. We see the two prominent
brightness edges that are the first cold fronts reported in
\citetalias{2000ApJ...541..542M} (marked ``southern'' and ``NW''). The
current, much deeper image reveals that the southern front spirals inward and
ends with another cold front (marked ``inner''). The inner front has been
noted by \cite{2011PhDT........14J} in the earlier \chandra\ dataset. A
temperature map of this structure (\autoref{fig:southern_inner}(a)) confirms
that the gas behind those brightness edges is cool, thus the cold front
interpretation is correct. A closer look at the image reveals that the
southern front branches in two, one branch apparently continuing with a
similar low curvature to the east (where we will find an intriguing
``channel'', \autoref{sec:channel}) and another one curving toward the center
and the inner front. Such a pattern is predicted by hydrodynamic simulations
of gas sloshing for the recently formed fronts (see, e.g.,
\citetalias{2006ApJ...650..102A} and their Figure 7, panels 1.8--2.1~Gyr, or
Figure 2 in \citealt{2015ApJ...798...90Z}). At this stage, the fronts do not
yet form a complete spiral pattern and still exhibit the remainders of the
Rayleigh-Taylor instability that gives rise to cold fronts with successively smaller radii (\citetalias{2006ApJ...650..102A}).

It is not clear whether the NW front and its more distant opposite
(\citealt{2013A&A...556A..44R}, outside this \chandra\ image) are part of the
same slosing pattern as the inner two or they are caused by another
disturbance. A closer look at \autoref{fig:splash} and the unsharp-masked
image in \autoref{fig:eddies}(b), as well as the gradient image in
\cite{2016MNRAS.461..684W} hints at subtle filamentary brightness
enhancements that start at the NW front and go inward, as if they were
extensions of the southern front. While \citeauthor{2016MNRAS.461..684W}
interpreted them as projected KH instability of the NW front, they may instead
be the structures surviving from the stage when the cool gas currently in the
core detached from the NW front and sank inward. However, this speculation
is beyond the statistical accuracy of the present dataset.

The gas density peak, which is right under the inner cold front and is the
location of the coolest gas (\autoref{fig:southern_inner}(a)), is offset by
$\approx 30$ kpc from the BCG, which is likely to be the center of the
gravitational potential. We will discuss this in
\autoref{sec:showinner}.

The southern cold front shows structure that resembles eddies of the KH
instability, predicted by hydrodynamic simulations with sufficient resolution.  The
NW front exhibits interesting structure consistent with such disturbances as
well. We will discuss the constraints on viscosity that we can place using
these observations in \autoref{sec:kh}. We start below with the necessary
preparatory analysis of the fronts.

\subsection{Southern Front}
\label{sec:show-southern}

We selected a sector enclosing the sharp segment of the southern cold front,
as shown in \autoref{fig:southern_inner}(b), and extracted a surface
brightness profile from the exposure-corrected image
(\autoref{fig:southern_inner}(c)) to model the 3D gas density across the
front. Our model describes the density profile inside the cold front with a
power law and outside the cold front with a beta model, with a density jump at
the cold front:
\begin{equation}
\label{eq:density}
  n(r) = \begin{cases}
    n_0 (r/r_J)^{\alpha} &,\; r \leq r_J \\
    \frac{n_0}{x} \left[ \frac{1+(r/r_c)^2}{1+(r_J/r_c)^2} \right]^{\beta}
                         &,\; r > r_J.
  \end{cases}
\end{equation}
Here, $r_J$ is the radius of the density jump, $x$\/ is the density jump
factor, $n_0$ is the density on the inside of the jump, and $r_c$\/ is the
core radius of the beta model. The model is centered at the center of
curvature of this section of the cold front (it is close to the X-ray
peak), and we assume spherical symmetry of the model (i.e. the same curvature
of the front along the l.o.s.\ as in the sky plane). The best-fit parameters
are given in \autoref{table:density-model-table}. The model fits the profile
very closely, showing a sharp jump at the cold front
(\autoref{fig:southern_inner}(c)).

We then extracted spectra from regions in the same sector on both sides of the
southern front and fitted their projected temperatures in XSPEC:
$T_{\text{cold,proj}}$ from a 10\arcsec\ wide annular segment inside, and
$T_{\text{hot,proj}}$ from a 15\arcsec\ wide annular segment outside, allowing
1\arcsec\ of clearance from the front position on either side. Using the APEC
normalization, we determined the absolute density by comparing it with the
model's emission measure $\int n_H n_e dV$, assuming $n_e = 1.17 n_H$.  To
evaluate the 3D gas temperature inside the cold front,
$T_{\text{cold,deproj}}$, we scaled the best-fit model in the outside region
by the ratio of our model's emission measure for the outside component that is
projected into the inner segment. We then refit the inner spectrum with this
component added and held constant. Finally, we used XSPEC to check if $x$\/
and $n_{\text{H,0}}$ needed to be corrected for the difference in 0.8--4~keV
emissivity in the presence of the temperature jump across the front (a small
factor not included in the brightness profile fitting procedure). For the
best-fit temperatures, the factor is $<$0.1\% so no correction was applied.
The temperatures are given in \autoref{table:density-model-table}. The gas
pressure across the front is continuous within the 90\% statistical
uncertainties.

\subsection{Gas velocity at the southern front}
\label{sec:vel}

For our instability analysis below, we now try to estimate the gas velocity at
the front. Within the simple subcluster-stripping picture of the fronts,
\citetalias{2000ApJ...541..542M} used the pressure profile to constrain the
velocity of the flow around the front, ascribing any difference of
thermodynamic pressures across the front to ram pressure. They obtained a
rough upper limit $v<400$~km~s$^{-1}$ for the southern front. A more accurate
way to estimate the front velocity from the pressure profile is proposed in
\cite{2001ApJ...551..160V} for A3667. However, we now think that (at least)
the southern and the inner fronts are, in fact, sloshing fronts with gas
flowing tangentially (see, e.g., \citetalias{2006ApJ...650..102A} for the
possible flow patterns). In particular, the cool gas under the southern front
is likely to be flowing from NW along the inward spiral.

In this picture, we can try to estimate the velocity of the curved tangential
flow from the centripetal acceleration, as was done in
\cite{2001ApJ...562L.153M} and \cite{2010ApJ...719L..74K}. In the simplest
approximation, the outer gas is stationary while the cold front gas inside the
front is in circular orbit with velocity $v$\/ in the cluster gravitational
potential. Then
\begin{equation}
\label{eq:centripetal}
  \frac{GM(r^{\prime})}{{r^{\prime}}^2} = - \frac{1}{\rho} \frac{dp}{dr} +
                                      \frac{v^2}{r^{\prime}},
\end{equation}
where $M$\/ is the cluster total mass within the radius $r^{\prime}$, $\rho$
is gas density, and $p$\/ is thermodynamic pressure. Here the $r$\/ coordinate
is from the center of the model density profile (the center of curvature of
the cold front) and $r^{\prime}$ is from the center of mass (the BCG). At the
cold front, they are at an angle of only 15$^{\circ}$, so
$dr^{\prime}/dr=0.97$ there, and we can ignore this distinction for an
approximate estimate. The left-hand side of \autoref{eq:centripetal} is
continuous over the cold front, because the cluster total mass distribution
(dominated by dark matter) is smooth. However, the moving gas inside the cold
front effectively feels a lower mass. Therefore, we can check for a difference
in the total mass derived under the hydrostatic equilibrium assumption (e.g.,
\citealt{1988xrec.book.....S}) on the inside and outside of the cold front,
and attribute it to the centripetal term. Using the gas density model of
\autoref{eq:density} and the temperatures on two sides derived above
(assumed constant at those values on both sides), we calculated the difference
between the 2nd term in \autoref{eq:centripetal} to be $(4.0\pm 2.7)\times
10^3$~km$^2$~s$^{-2}$~kpc$^{-1}$, which corresponds to a $\approx$35\% drop in
the apparent total mass on the inside of the cold front. The hydrostatic mass
given by the outer part of the model (i.e., the true mass under our
assumptions) is $(1.5\pm 0.3) \times10^{13}$~\msun\ within
$r^{\prime}=75$~kpc of the BCG.

This gives a tangential velocity of the cold gas of $(550 \pm
190)$~km~s$^{-1}$, where the errors are statistical and include the
uncertainties of the parameters $\alpha$, $\beta$, $r_J$, $r_c$,
$T_{\text{cold,deproj}}$, and $T_{\text{hot}}$
(\autoref{table:density-model-table}). If we use a smaller radius of curvature
such as that of the cold front at this position, we get a lower value but not
by much, because of the square root. Given the unknown 3D geometry and a
number of assumptions, this is, of course, only a qualitative estimate with a
factor 2 accuracy at best. Furthermore, if the outer gas also rotates but in
the opposite direction, one can in principle have a much higher {\em
relative}\/ tangential velocity and still satisfy \autoref{eq:centripetal}.
While such a scenario is unlikely considering how the sloshing fronts form ---
we do not expect flows faster than Mach $\sim$0.3--0.5 --- a conservative upper
limit on the relative velocity is probaby the sound speed in the outer gas
(1500~km s$^{-1}$), from the fact that we do not see any shocks immediately
outside this cold front. We will use the velocity estimate of 550~km~s$^{-1}$ in
\autoref{sec:kh} below.

\begin{table*}[th!]
\centering
\caption{Best fit cold front model parameters. $n_{H,0}$ is given as the model
density on the inside of the jump, calculated using density and temperature of
the outer component. Errors are 90\%. \label{table:density-model-table}}
\begin{tabular}{llllllllll}
\noalign{\smallskip}
\toprule
Location & $n_{H,0}$   & $r_J$ & $x$ & $\alpha$ & $\beta$ & $r_c$ &
    $T_{\mathrm{cold, proj}}$ & $T_{\mathrm{cold, deproj}}$ & $T_{\mathrm{hot}}$\\
\, & $10^{-3}$\,cm$^{-3}$ & kpc & \, & \, & \, & kpc & keV & keV & keV \\
\noalign{\smallskip}
\hline
\noalign{\smallskip}
Southern & $16.4\pm0.16$ & $63.9^{+0.6}_{-0.5}$ & $1.87\pm0.1$ &
    $-0.51\pm0.09$ & $-0.60^{+0.04}_{-0.05}$ & $75^{+16}_{-15}$ &
    $6.9^{+0.8}_{-0.5}$ & $5.8^{+1.1}_{-0.9}$ & $9.0^{+1.1}_{-0.9}$ \\
\noalign{\smallskip}
NW1 & $4.31\pm0.04$ & $175.0\pm1.0$ & $2.14^{+0.09}_{-0.10}$ &
    $-0.42\pm0.04$ & $-0.71^{+0.05}_{-0.06}$ & $218^{+39}_{-37}$ &
    $8.6^{+1.2}_{-0.8}$ & $7.9^{+1.6}_{-1.3}$ & $10.5^{+1.9}_{-1.2}$ \\
\noalign{\smallskip}
NW2 & $4.31\pm0.04$ & $174.6^{+1.0}_{-1.7}$ & $2.07^{+0.11}_{-0.10}$ &
    $-0.50\pm0.04$ & $-0.66\pm0.05$ & $178^{+41}_{-43}$ &
    $7.2^{+0.9}_{-0.7}$ & $6.1^{+1.1}_{-0.9}$ & $10.5^{+2.2}_{-1.3}$ \\
\noalign{\smallskip}
\hline
~\\
\end{tabular}
\end{table*}

\subsection{The Displaced Gas Peak and the Inner Front}
\label{sec:showinner}

We noted above that the gas density peak is offset $\sim$30~kpc from the
position of the brightest cluster galaxy (\autoref{fig:splash},
\autoref{fig:southern_inner}(b)). Such offsets are rare but not unknown ---
they have been seen, e.g., in A644 \citep{2005ApJ...630..750B}, Ophiuchus
(\citealt{2010MNRAS.405.1624M}; \citealt{2010ApJ...717..908Z};
\citealt{2012MNRAS.421.3409H}; \citealt{2016MNRAS.460.2752W}), A1991
\citep{2012MNRAS.421.3409H}, and Zw1742+33 \citep{2013A&A...555A..93E}.
However, to our knowledge, this is the largest offset seen in a cluster that
still has a cool density peak. A comparable offset of 20~kpc is seen in
Zw1742+33, but that cluster also shows evidence of AGN X-ray cavities
emanating from its BCG, which has an active nucleus seen in the radio and
X-rays. In comparison, the BCG in A2142 is currently very faint in the radio
\citep{2017A&A...603A.125V} and is not detected in the X-ray; we see no
evidence for X-ray cavities either. Thus, the offset peak that we observe is
clearly the result of sloshing and of the merger that set it off.

\begin{figure*}[!ht]
\label{fig:bcg_displace}
        \centering
        \leavevmode
        \includegraphics[width=85mm]{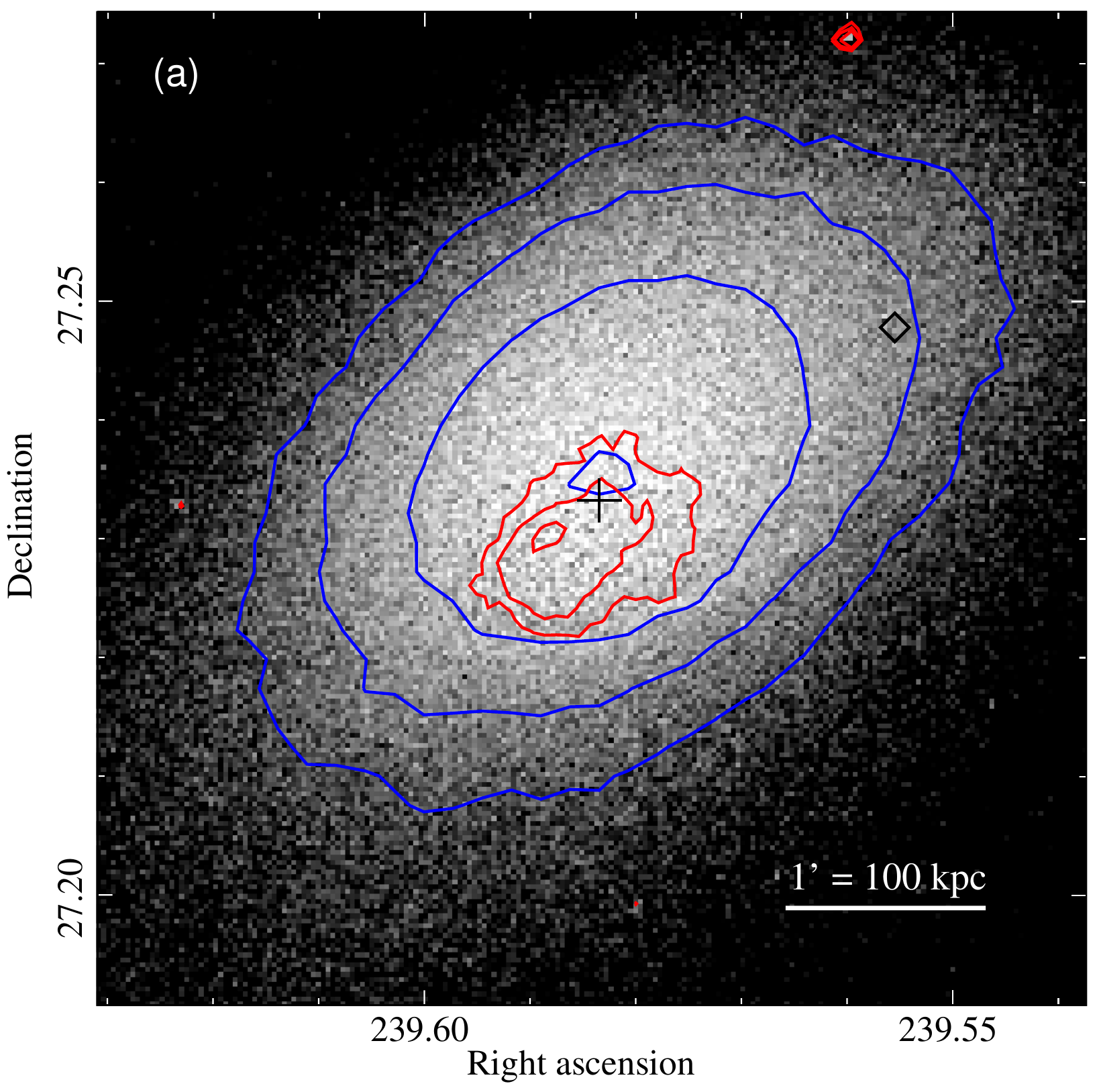}
        \hfil
        \includegraphics[width=85mm]{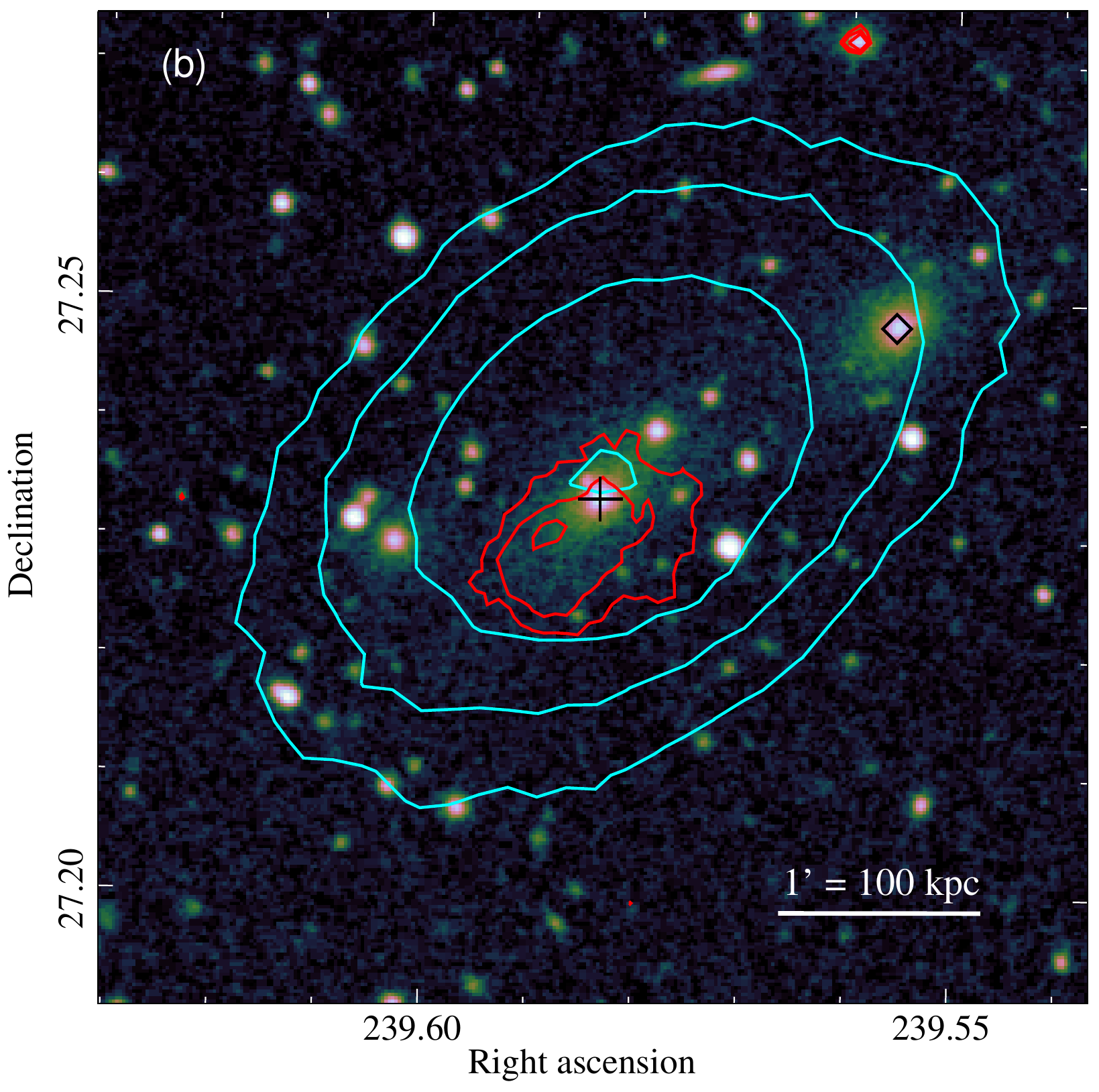}

        \caption{
          (a) Residual 0.8--4~keV flux, binned to 1.5\arcsec\ pixels, after
          subtracting wavelet components 20~kpc and below.
          (b) Optical image from Digital Sky Survey image archive, showing the
          main BCG (marked by the cross) and its neighborhood. The position of
          the second brightest galaxy is also shown (marked by the diamond).
          The blue contours show the position of the peak and the shape of the
          residual emission in (a), with levels in 1.4$\times$ steps.
          The red contours show the X-ray peak and shape of the small-scale
          structures, with levels in 2$\times$ steps. They are derived from a
          wavelet reconstruction of the small scale structures using scales up
          to 39~kpc.
          The wavelet reconstruction uses the same decomposition as described
          in \autoref{sec:temperaturemap}.}
\end{figure*}

We subtracted the cool sloshing structure from the X-ray image by wavelet
decomposition to see the larger-scale X-ray gas distribution. After the
subtraction of components 20~kpc and smaller (using the same decomposition as
\autoref{sec:temperaturemap}), we are left with the image shown in
\autoref{fig:bcg_displace}(a). The contours of the subtracted small-scale
structure are overlaid. We see a symmetric elliptical X-ray structure centered
very near the BCG. This is consistent with a picture where the BCG is the
center of the gravitational potential of the cluster, and the gas beyond the
inner sloshing structure is largely in hydrostatic equilibrium with it (this
does not exclude slower motions that can accompany the outer cold fronts). The
gravitational lensing map of \cite{2008PASJ...60..345O} does show the main
mass peak of the cluster near this BCG. The second brightest galaxy seen in
\autoref{fig:bcg_displace}(b), which was though in
\citetalias{2000ApJ...541..542M} to be the center of a merging subcluster,
appears not to be physically related to the cluster, based on its high
peculiar velocity (1840~km~s$^{-1}$ from the BCG,
\citealt{1995AJ....110...32O}) and lack of a mass concentration
\citep{2008PASJ...60..345O}.

We will now model the inner cold front in order to derive the parameters of
the gas in the offset density peak. A surface brightness edge near the peak of
the X-ray emission spans a sector from east to north
(\autoref{fig:southern_inner}(b)). The contrast in X-ray brightness and
projected temperature is highest in the northeastern quadrant, and the edge
disappears to the west. It is a cold front, as shown by the temperature map
(\autoref{fig:southern_inner}(a)). We extracted a brightness profile
(\autoref{fig:southern_inner}(d)) in the sector show in
\autoref{fig:southern_inner}(b) and model it it as follows. The density
profile inside the edge is centered on the center of curvature of the front
and is a power law. The outer gas is modeled with an ellipsoidal component
following a power law profile, centered on the BCG. The ellipticity of the
outer component model is achieved simply by rescaling the coordinate of the
long axis before calculating the model density in 3D. Both position angle and
ellipticity of the outer component were deduced from the X-ray brightness
contours of the remaining cluster emission after we subtracted the core
structure (as described above), and fixed during the fit. Since the two
density components have different centers, we could not just calculate a 1D
projected model. Instead, we projected the model onto the same image plane as
the flux image and extracted a brightness profile in the same sector. The best
fit model is shown in \autoref{fig:southern_inner}(d). To determine a
deprojected central temperature, we first fitted the spectrum extracted from a
sector, 17~kpc wide, just outside the cold front. Then, we created an image
of the ellipsoidal component with a spherical cutout for the core and used it
to normalize the projected contribution to an inner sector, 10~kpc thick,
inside the front. We then fixed this contribution at the best-fit outer
temperature and fit the inner temperature. Finally, we use the APEC model
normalization to derive the gas densities in 3D as we did in
\autoref{sec:show-southern}.

Our deprojected density just behind the cold front (near the peak) is $n_H
\approx 2.3\times10^{-2}$~cm$^{-3}$ and temperature $T=4.0^{+0.8}_{-0.6}$~keV.
The gas specifc entropy index, commonly defined in the cluster field as $K = T_e
n_e^{-2/3}$, is $K\approx49$~keV~cm$^2$ (statistical errors are probably
meaningless because the systematic uncertainties dominate). The true value at
the peak can be slightly lower because our spectral fitting region does not
resolve the peak. For the gas immediately outside this cold front, our model
gives $K\approx120$~keV~cm$^2$. We note that our value for the central entropy index
is lower than $58\pm2$~keV~cm$^2$ in \cite{2017ApJ...841...71G} from the same
dataset; however, the difference is expected because those authors have used a
different definition of ``central entropy'' in order to be consistent with
\cite{2009ApJS..182...12C}, who combined the projected temperature with the 3D
gas density, whereas both our quantities are deprojected.

The above small difference notwithstanding, our value of the central entropy
places A2142 in the gap between the cool-core and non-cool-core clusters
\citep{2009ApJS..182...12C}. This is apparently related to strong sloshing in
this cluster. As shown by \cite{2010ApJ...717..908Z}, sloshing of a cool core
can balance radiative cooling, except for the very central region, by
facilitating mixing with the higher-entropy gas from outside the core. Once
the gas peak is displaced from the minimum of the gravitational potential, it
becomes even more prone to mixing, because it expands (which reduces the
density contrast) and because the stabilizing effect of gravity is removed. We
may have caught A2142 at the moment of dissolution of its former cool core by
sloshing. The displacement of the gas peak should also have deprived the cD
galaxy of the accreting cool gas for a significant period of time, which is
why it does not exhibit an AGN, similarly to Ophiuchus and to most clusters
without cool cores.

\subsection{NW Front}
\label{sec:nwcoldfront}

\begin{figure*}[!ht]
\label{fig:northwest_profile}
        \centering
        \leavevmode
        \gridline{
        \fig{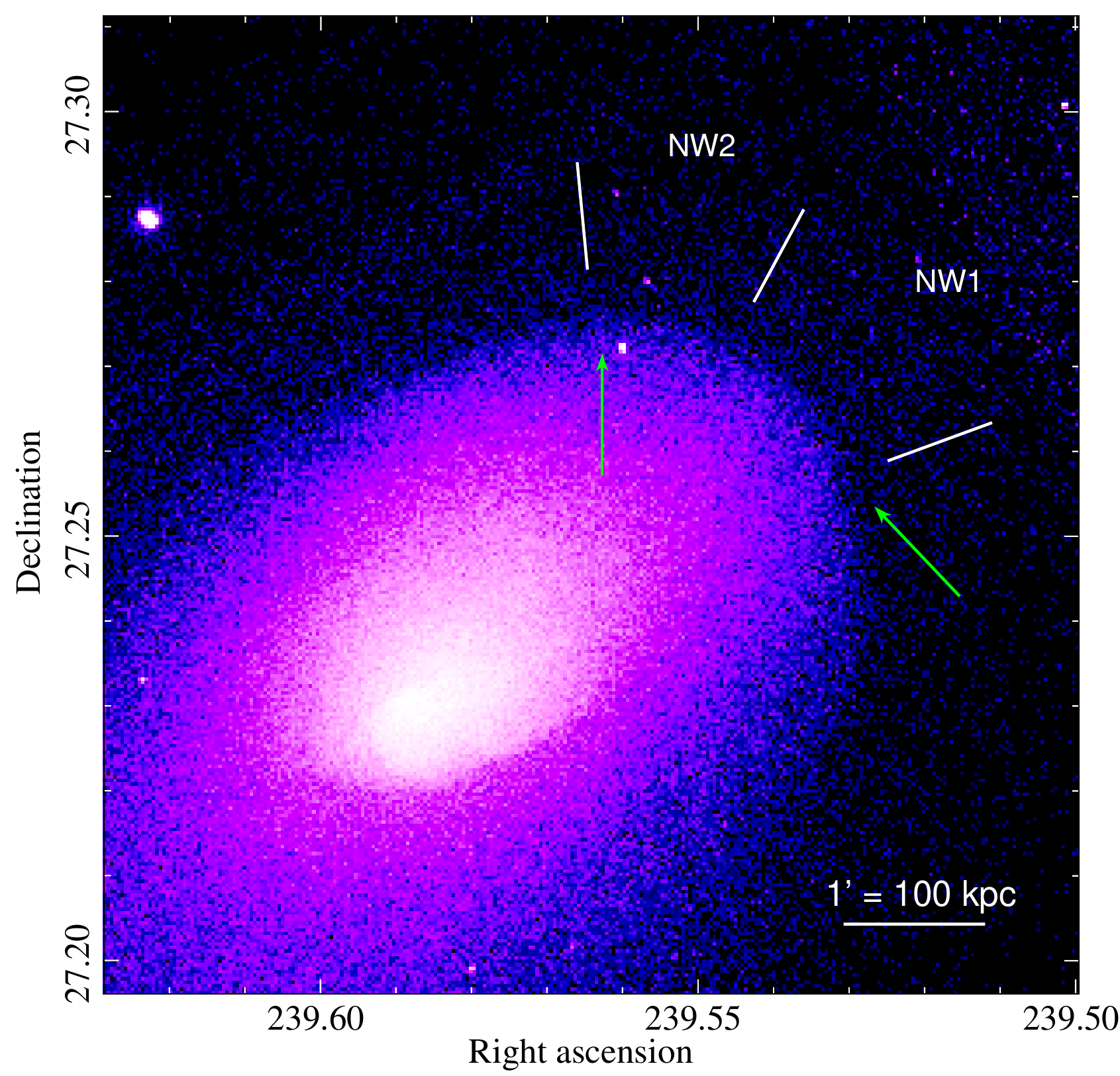}{0.5\textwidth}{(a)}
        \fig{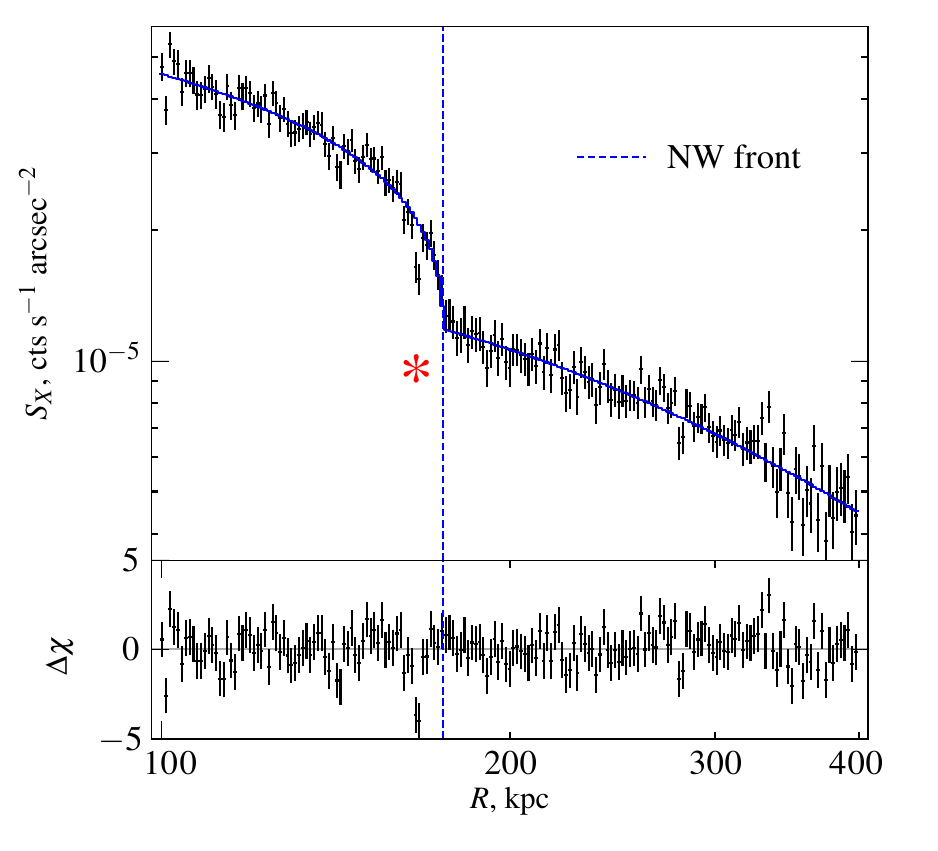}{0.5\textwidth}{(b)}
        }
        \caption{
          (a) The NW cold front, with the brightness profile sectors NW1 and
          NW2 marked, on a 0.8--4~keV image binned to 1.5\arcsec\ pixels. The
          green arrow to the left points to the ``boxy'' shape of the front.
          The continuation of the right green arrow is the feature that shows
          as a dip in surface brightness.
          (b) Brightness profile in the NW1 sector. There is a highly significant
          drop in X-ray brightness, at the radius indicated by the red asterisk, 8--10~kpc inside the best-fit
          position of the density jump (dashed line). The best-fit positions are statistically
          identical in the two sectors.}
\end{figure*}

Upon close inspection, the NW front (\autoref{fig:northwest_profile}) shows
interesting structure, which includes a ``boxy'' shape and apparent multiple
edges at its nose. We extract brightness profiles in sectors NW1 and NW2 shown
in \autoref{fig:northwest_profile}(a) and fit them as in
\autoref{sec:show-southern} with the density model given in
\autoref{eq:density}, centered on the front center of curvature (same for both
sectors). The best-fit parameters, along with the gas temperatures across the
front, are given in \autoref{table:density-model-table}. For the observed
temperature jumps, a 1\% reduction was applied to the jump factor to correct
for the higher 0.8--4~keV emissivity at the lower deprojected temperature.
These two segments of the cold front are visually similar, have the same
radius of curvature, and the brightness jump can be traced by the same circle.
Their model parameters are therefore very comparable, and indeed their
best-fit density jump positions, jump factors, and outer model index $\beta$
are consistent with being the same. The inner index $\alpha$ and the beta
model core radius $r_c$ are statistically different, but this is expected
because the cluster's ellipticity. The brightness profile and the the best-fit
model for NW1 are shown in \autoref{fig:northwest_profile}(b) (the NW2 profile
is not shown as the fit is good and there is nothing special about it.)
Notably, the NW1 brightness profile shows a $4\sigma$ dip --- 25\% below the
model --- 8--10~kpc behind the front. This feature is seen in the image in
\autoref{fig:northwest_profile}(a) (on the continuation of the right arrow).
Along with the boxy shape (left arrow), it looks just like the deformations
expected from KH instabilities (e.g., \citealt{2013ApJ...764...60R}, see their
Figure 6) and seen in a few other clusters. In particular, multiple edges would
be the KH eddies that develop along the line of sight.

\section{Constraints on plasma viscosity}
\label{sec:kh}

\begin{figure*}[!ht]
        \centering
        \leavevmode
        \includegraphics[width=85mm]
                {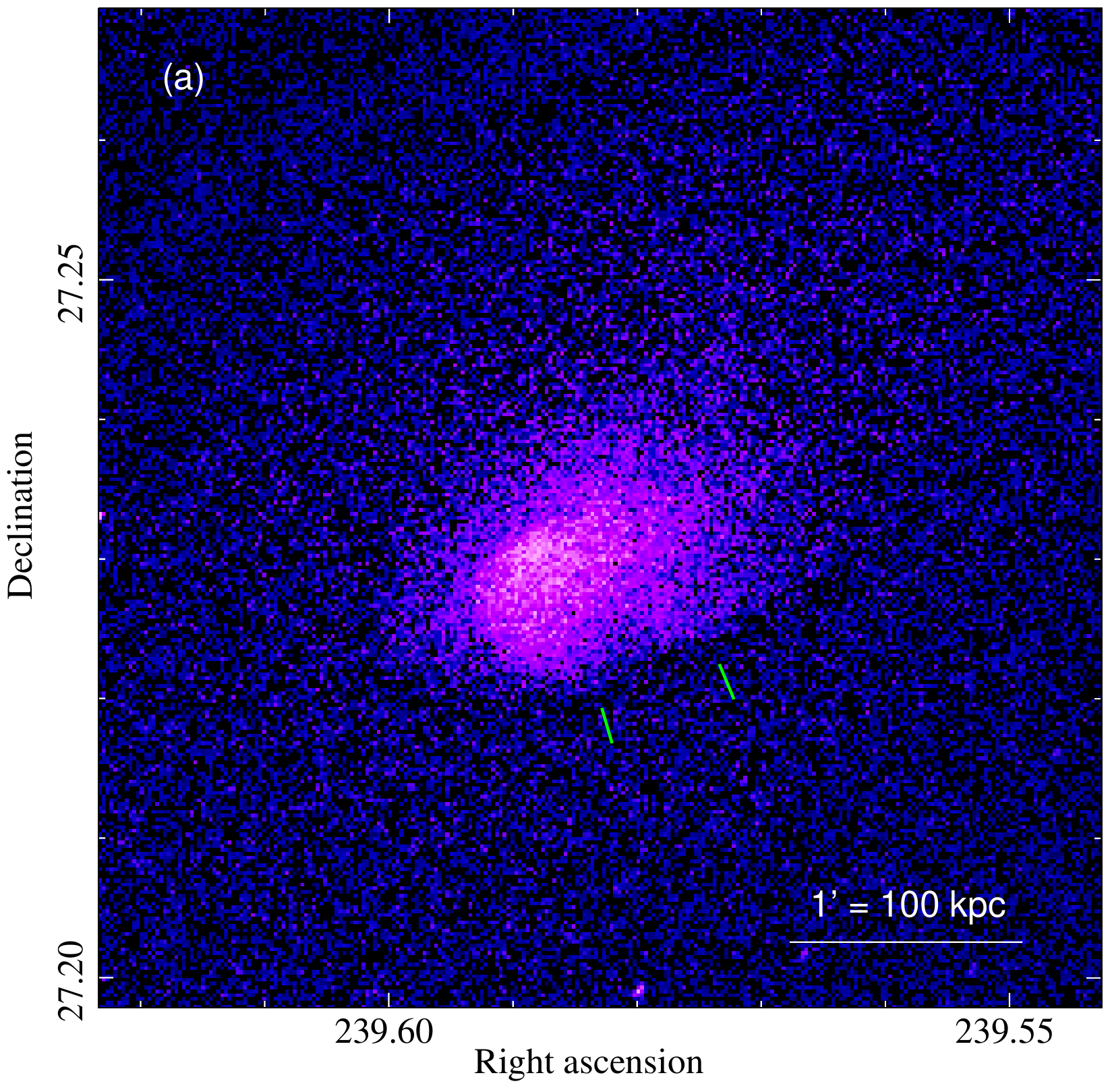}
        \hfil
        \includegraphics[width=85mm]
                {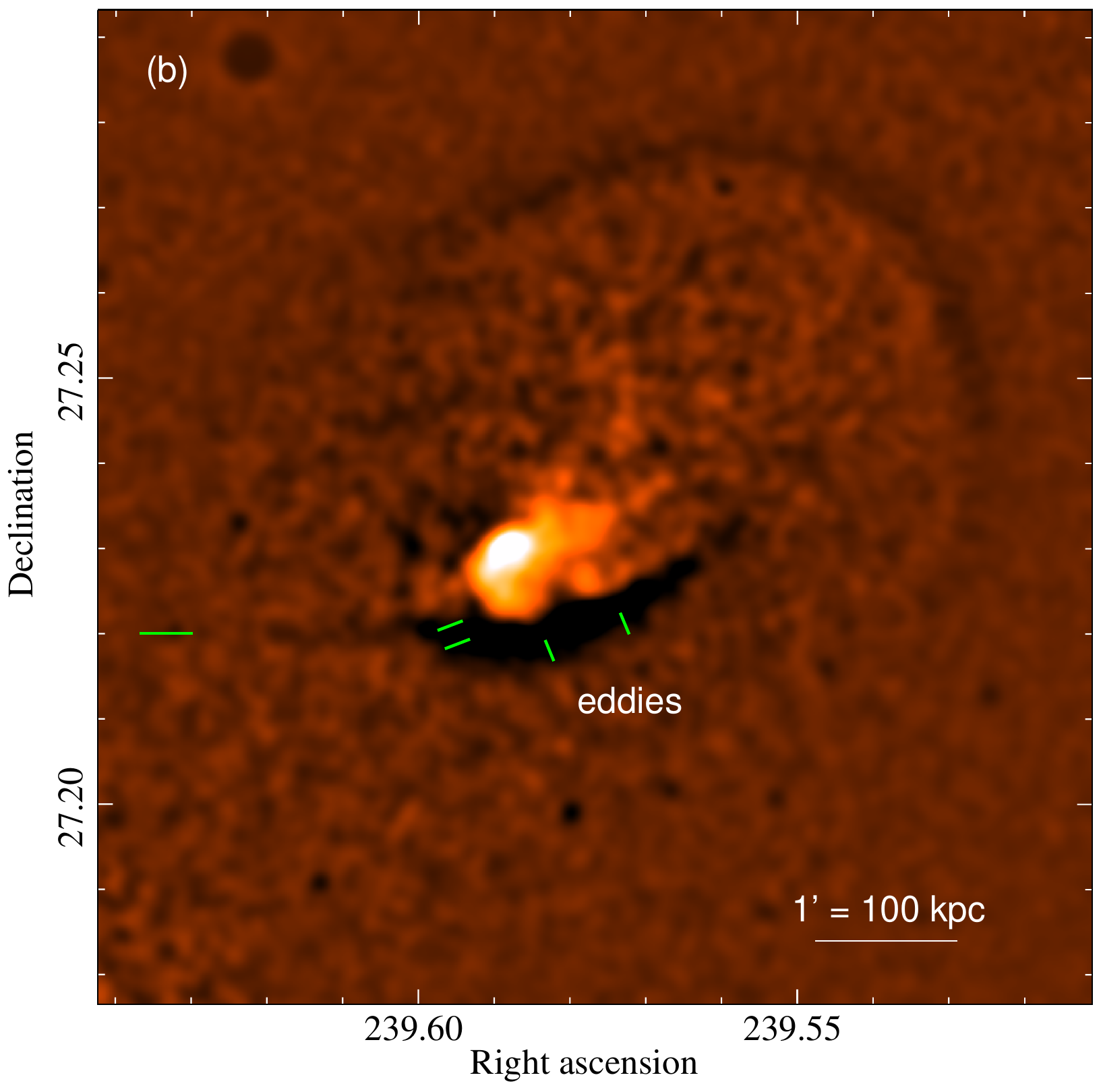}

        \caption{
          Zooming in on the suspected Kelvin-Helmholtz eddies at the southern
          cold front.
          (a) Wavelet decomposition was used to remove emission from
          components on scales larger than the KH eddies, 16\arcsec\ (=26~kpc) and
          up, by subtracting them from the 0.8--4~keV image binned to 1\arcsec\
          pixels. The two green ticks mark the crests of the KH eddies.
          (b) Unsharpmasked version of the 0.8--4 keV image, created by
          subtracting one image smoothed by a $\sigma$=12\arcsec\ gaussian kernel
          from a second image smoothed by $\sigma$=3\arcsec, so as to highlight
          features on scales in between. The additional pair of green ticks
          mark the crest-to-trough scale of the eddies. The horizontal green
          line to the left points along the channel discussed in
          \autoref{sec:channel}.
          }
        \label{fig:eddies}
\end{figure*}

Even without any image enhancements, the X-ray image of the southern front
(Figures \ref{fig:splash}, \ref{fig:southern_inner}(b)) shows a wavy structure
that looks like the classic KH instability (KHI) at the interface of two gas
layers with velocity shear. In \autoref{fig:eddies}(a), we show a slightly
enhanced image of the small-scale structure by subtracting the large-scale
($\geq$26~kpc) wavelet components from the raw image. In
\autoref{fig:eddies}(b), we instead apply the usual unsharp mask. Both images
reveal two prominent bumps of the cold front surface that we interpret as two
developed KH eddies, spaced by 55~kpc, with a crest-to-trough amplitude of
13--15~kpc (green dashes in \autoref{fig:eddies}(b)). This amplitude is a
lower limit because projection can only make it look smaller. The high
contrast of the edge suggests that we are getting an edge-on view of the shear
layer. This is only the second cold front that affords us a good, direct, and unambiguous view
of the KH eddies; the other one is A3667 (\citealt{2011scgg.conf...53V};
\citealt{2017MNRAS.467.3662I}).

If these are indeed KH eddies, they present an opportunity to constrain the
ICM effective viscosity. In our picture, the gas inside the southern cold
front is flowing along the curved edge from the NW and spirals inward with the
velocity that we estimated in \autoref{sec:vel}, while the outer gas has a
negligible velocity. \citetalias{2013MNRAS.436.1721R} performed a numerical
study of the growth of KH instabilities on cluster cold fronts for a range of
values of isotropic viscosity (under the assumption of no magnetic fields)
both Spitzer-like with strong temperature dependence as well as
temperature-independent. They covered a range of gas parameters that included
the A2142 southern front --- in fact, they used it as one of their fiducial
cases (using the early \citetalias{2000ApJ...541..542M} results that did not
show the eddies). While the \citetalias{2013MNRAS.436.1721R} simulations are
2D, they should provide a good qualitative approximation for the flow geometry
expected at the cold front. Thus, all we need is to find where our new results
fit in the \citetalias{2013MNRAS.436.1721R} study to derive an estimate of the
viscosity under their assumptions. We will try to constrain the isotropic
Spitzer-like viscosity.

A full Spitzer viscosity would suppress the growth of KHI on small scales, so
that only the perturbations of the interface between the two fluids larger
than a critical wavelength can grow (\citetalias{2013MNRAS.436.1721R}, their
Eq.\ 28):
\begin{equation}
\label{eq:lambda_crit}
\begin{split}
  \lambda_{\mathrm{crit}} = \;70\,\mathrm{kpc}\,
  &\left( \frac{\mathrm{Re_{crit}}}{30} \right)
  \left( \frac{U}{500\,\mathrm{km\,s}^{-1}} \right)^{-1} \times \\
  &\left( \frac{n_e}{9 \times 10^{-3}\,\mathrm{cm}^{-3}} \right)^{-1}
  \left( \frac{kT_{\mathrm{ICM}}}{9.0\,\mathrm{keV}} \right)^{5/2},
\end{split}
\end{equation}
where the density and temperature are those observed on the hotter side of the
front from \autoref{table:density-model-table} (because the temperature
dependence of the Spitzer viscosity makes that side dominate the effect), $U$
is the relative shear velocity of the gases on two sides of the cold front and
$\mathrm{Re_{crit}}$ is a Reynolds number defined for the KHI as
\begin{equation}
\label{eq:KH_Re}
  \mathrm{Re} \equiv \frac{\lambda U}{\nu},
\end{equation}
where $\nu$ is the kinematic viscosity. The full Spitzer viscosity is
\citep{1962pfig.book.....S, 1988xrec.book.....S}
\begin{equation}
\label{eq:spitzer_viscosity}
  \mu = 6100\;\mathrm{g \;cm^{-1}\, s^{-1}}\;
  \left(\frac{kT}{9.0\,\mathrm{keV}}\right)^{5/2}
  \left(\frac{\ln \Lambda}{40}\right)^{-1} ,
\end{equation}
where $\nu = \mu / \rho$ and $\ln \Lambda \approx 37$ for the density and
temperature we measure outside the southern cold front.

Based on simulations, \citetalias{2013MNRAS.436.1721R} showed that for
Spitzer-like viscosity, a conservative value is $\mathrm{Re_{crit}}=30$ to
suppress KHI. We do see a developed KH instability, so for our wavelength,
$\mathrm{Re}>30$. To place a somewhat more accurate lower limit on the
Reynolds number, and thus an upper limit on the viscosity, we compare our
eddies with those in the \citetalias{2013MNRAS.436.1721R} simulations at a
similar growth stage. Figure 8 in \citetalias{2013MNRAS.436.1721R} shows the
development of KHI for different Reynolds numbers and the interface parameters
very close to ours (their density contrast is 2 vs.\ our 1.9 and their $M=0.5$
vs.\ our rough estimate of $0.36\pm 0.12$). For our front, we can use the
peak-to-peak distance to measure the KHI $\lambda \simeq55$~kpc. The amplitude
(half of the crest-to-trough distance) appears to be at least
0.10--0.12$\lambda$.  There are not enough photons to resolve the small-scale
features in the eddies, such as the expected turning-over of the tip of the
eddy, though observers with imagination would see a hint of this in the
wavelet-subtracted image.

We can estimate the time that the eddies had to grow to their present
amplitude. The inviscid KH timescale (\citetalias{2013MNRAS.436.1721R}, their
Equations 2-3) is
\begin{equation}
\tau_{\mathrm{KHinvis}} = \frac{\sqrt{\Delta}}{2 \pi} \frac{\lambda}{U} ,
\end{equation}
where
\begin{equation}
  \Delta = \frac{(\rho_{\mathrm{cold}} + \rho_{\mathrm{hot}})^2}
                {\rho_{\mathrm{cold}}\rho_{\mathrm{hot}}}
\end{equation}
is related to the growth time of the eddies estimated from $t=L/U$, by
\begin{equation}
\frac{t}{\tau_\mathrm{KHinvisc}} = \frac{2\pi L}{\sqrt{\Delta} \lambda}.
\end{equation}
If we take the distance $L$\/ that the perturbations have traveled along the
front to be from the crests of the eddies to the eastern side of the front,
$L\approx50$--$100$~kpc and $t\approx 3$--$6\tau_\mathrm{KHinvisc}$.

If we compare our eddy amplitude to \citetalias{2013MNRAS.436.1721R} at this
early growth stage (see their Figure 8 and the left panel of Figure 10), they
look similar to the case with $\mathrm{Re}=100$ or above and rule out Reynolds
numbers much lower than that. We note that the
\citetalias{2013MNRAS.436.1721R} simulations assumed uniform density on each
side of the interface, whereas our density increases toward the cluster center
(away from the interface) and changes noticeably on the scale of the
disturbance. This is likely to decrease the depth of the troughs compared to
the simulated case, so the above estimate should be conservative.

To convert this to a constraint on the viscosity, \autoref{eq:KH_Re} and
\autoref{eq:spitzer_viscosity} can be combined:
\begin{equation}
\begin{split}
  \frac{\mu_S}{\mu} = 5
  \left(\frac{n_H}{9 \times 10^{-3} \, \mathrm{cm}^{-3} }\right)^{-1}
  \left(\frac{kT_e}{9.0\,\mathrm{keV}}\right)^{5/2}
  \left(\frac{\ln \Lambda}{40}\right)^{-1}\\
  \times
  \left(\frac{\mathrm{Re}}{100}\right)
  \left(\frac{\lambda}{55\,\mathrm{kpc}}\right)^{-1}
  \left(\frac{U}{500\,\mathrm{km\,s}^{-1}}\right)^{-1}.
\end{split}
\end{equation}
Here we again used the values of the gas density and temperature on the hotter
side of the front, the shear velocity that we estimated in
\autoref{sec:vel}, and the above wavelength and Reynolds number of the KHI.
The velocity of the flow is the most uncertain parameter for our constraint,
but even if we use a very conservative upper limit of 1500~km~s$^{-1}$
(\autoref{sec:vel}), the viscosity should still be lower than Spitzer.

The NW front also shows hints of KH instabilities, including the boxy shape of
the front and the apparent double density edge seen in projection
(\autoref{sec:nwcoldfront}). They are not seen directly in the plane of the sky
as the southern front eddies, so any constraints from them would be more
uncertain than those above. However, the NW edge samples a factor 4 different
gas density and possibly a different velocity, so it may be interesting to
perform a joint study of the two edges, perhaps using hydrodynamic simulations
to reproduce their morphology and better constrain the flow velocities.

The \citetalias{2013MNRAS.436.1721R} simulations have a major omission ---
they do not include magnetic fields, which we know are present in the
intracluster plasma and, furthermore, should be significantly amplified and
stretched along the cold front surface because of the expected draping
(\citealt{2000MNRAS.317L..57E}; \citealt{2001ApJ...551..160V};
\citealt{2006MNRAS.373...73L}; \citealt{2008ApJ...677..993D}).
\cite{2015ApJ...798...90Z} showed via high-resolution MHD simulations that in
the context of sloshing cold fronts in clusters, isotropic Spitzer viscosity
reduced by a factor $\sim$0.1 produces similar-looking cold fronts as the
anisotropic Braginskii viscosity that describes the magnetized plasma. Thus,
our estimate of the effective viscosity of $<$$1/5$ Spitzer is in agreement
with full anisotropic viscosity in the presence of the magnetic fields.

\section{X-ray Channel}
\label{sec:channel}

\begin{figure*}[!ht]
        \centering
        \leavevmode
        \gridline{
        \fig{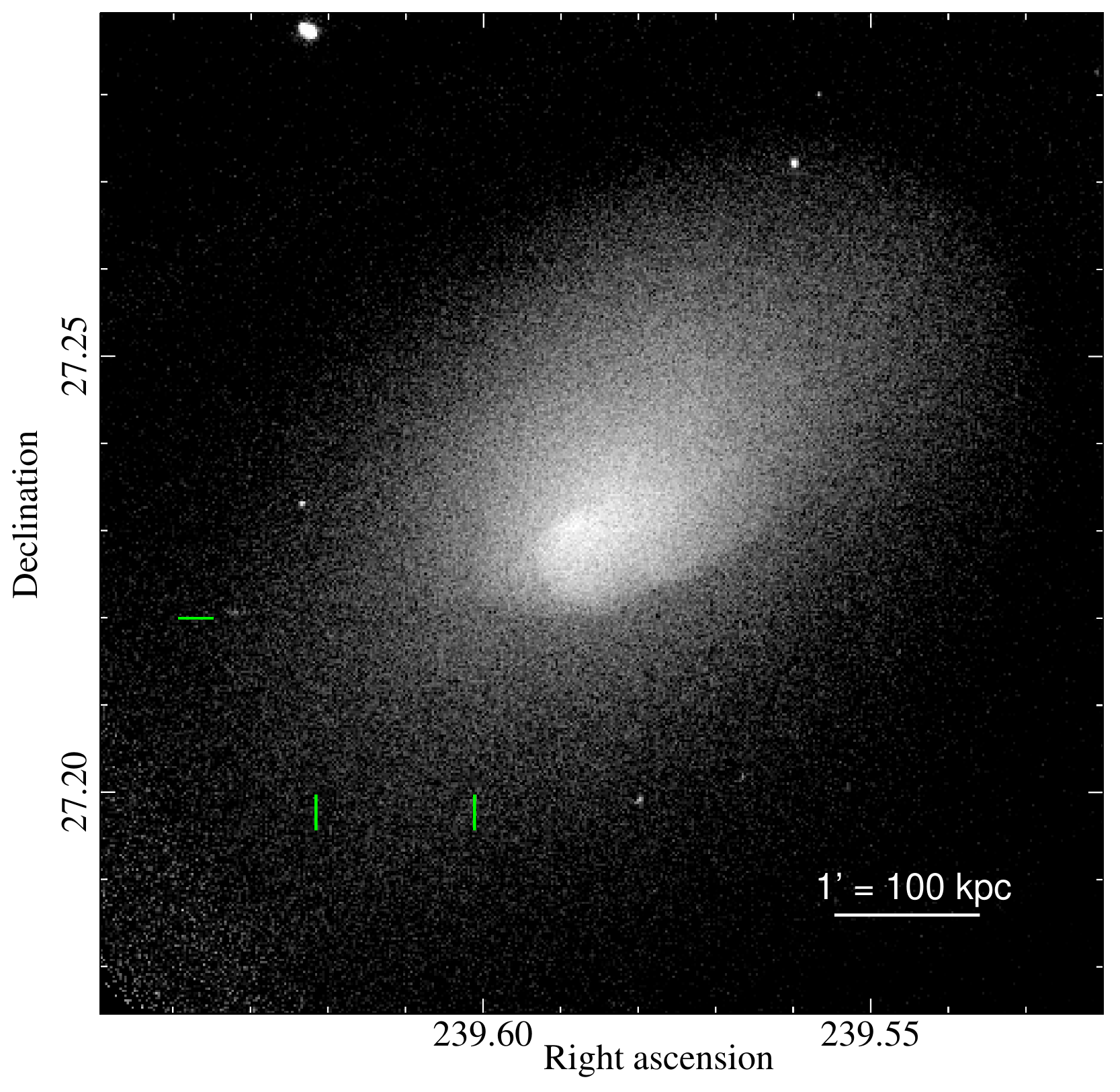}{0.5\textwidth}{(a)}
        \fig{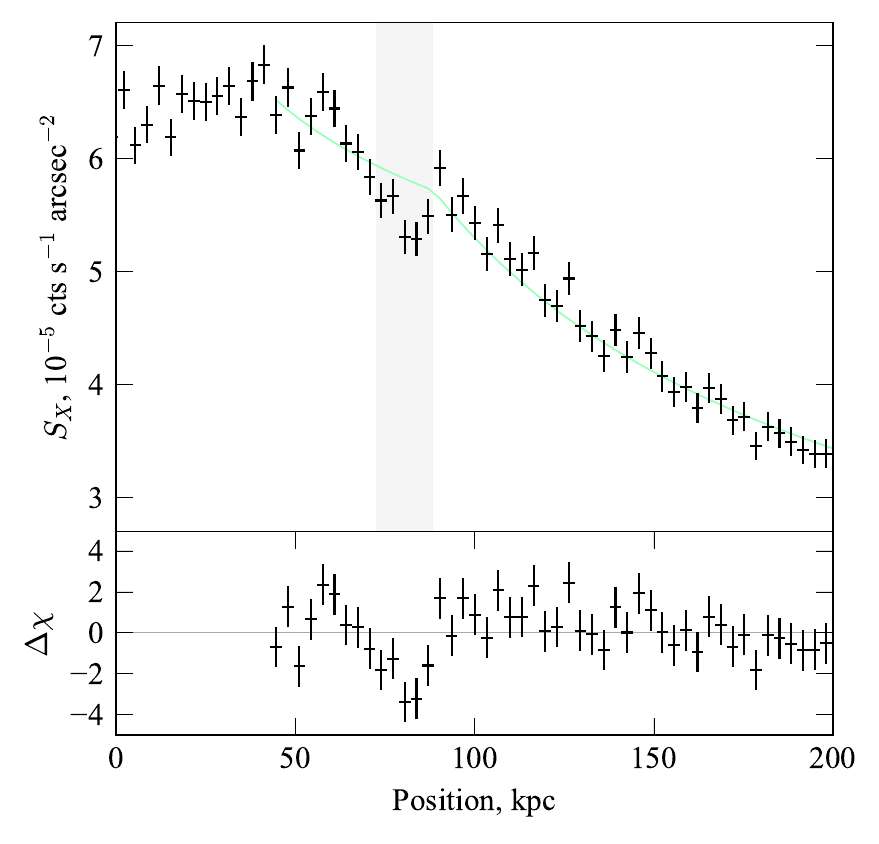}{0.5\textwidth}{(b)}
        }
        \caption{
          (a) Image in the 0.8--4.0~keV band, binned by 1\arcsec, with colors
          selected to better show the channel. The green horizontal tick
          shows the position of the channel, which can also be seen in the
          unsharp-masked image in \autoref{fig:eddies}(b). The vertical ticks
          mark the span of the rectangular band used to derive the brightness
          profile shown in panel (b).
          The position coordinate in the profile runs from north to south
          (zero is arbitrary). Error bars for surface brightness are
          1$\sigma$.
          The shaded band indicates the apparent width of the channel.
           The green line shows a simple best-fit generic model that would
          represent a break (but no dip or a jump up) in the density profile,
          and residuals in the lower panel are for this model.}
\label{fig:channel_profile}
\end{figure*}

There is a subtle, long X-ray brightness channel that extends from the the
middle of the southern cold front to the east. We selected the contrast
in the unbinned X-ray image shown in \autoref{fig:channel_profile}(a) to
emphasize this linear depression. The unsharp-masked image in
\autoref{fig:eddies}(b) helps to see the feature's location. It is not a
residual artifact of any ACIS chip gaps or edges, which are corrected for in
all our images. Further, in this mosaic of slightly different pointings, the
feature does not overlap with any chip gap or edge. Even if the exposure maps
were significantly inaccurate, the amplitude of the effective exposure
variations over the channel region that it corrects for is $<$2\%, while the
depth of the channel is much greater. The channel is aligned with the
southern cold front (with its branch that does not curve toward the center but
continues eastward, \autoref{sec:thecoldfronts}). While the channel is most
apparent to the east of the front, it may continue west, wrapping around the
southern front. However, the much greater brightness gradient associated with
the front there, as well as the KH eddies, preclude the detection of a subtle
dip, because the baseline brightness is very uncertain.

While large, apparently significant deviations, some arranged in patterns, are
expected in a noisy image with many independent pixels (the ``look elsewhere''
effect), this apparent linear feature is not found at a random place, but
rather at a continuation of a prominent cold front. So it is likely to be a
real structure.

We selected a section of the channel 110~kpc long, where the channel is
unobstructed by brighter features, and extracted a brightness profile across
it in a strip indicated by the tick marks in \autoref{fig:channel_profile}(a).
The width of the channel is about 15~kpc. The brightness profile is shown in
\autoref{fig:channel_profile}(b), where each bin is 3.3~kpc (2\arcsec) wide.
To quantify the dip amplitude, we performed a simple fit of the
brightness in the vicinity of the dip with a broken power-law model (which
would represent a break in the density profile, but not allowing for a dip or
a density jump up or down at the break), shown in green. There is a very
significant $\sim$9--12\% depression in the surface brightness at the center
of the channel, where two bins are each $>$3$\sigma$ below the
model and below the brightness in bins immediately to the right (outwards).

The origin of such a density depression is not immediately clear. Simply
projecting any number of monotonically declining brightness profiles of any
shape would not create a brightness depression (but could create multiple
brightness edges, as seen elsewhere in A2142) --- as long as the density
gradients point in the same general direction of the cluster center. One can
imagine two cold fronts {\em facing}\/ each other, with their gradients in the
opposite directions, as in two subcluster cores about to collide and a
low-density layer between them. However, based on the X-ray image, such a
scenario is clearly not the case in A2142. Perhaps some other unexpected gas
geometries could emerge in a merging cluster.

If we take the premise that the feature is indeed due to a density
depression, not the presence of an edge-like profile facing the opposite
direction, the geometry of this channel has to be a relatively thin sheet of
lower-density gas, seen along its edge. If we consider the surface brightness
profile of a NE-SW cross section of the cluster at the position of the
channel, we must empty of gas the central 35~kpc interval along the l.o.s.\ to
remove 10\% of the flux.  Since the channel cannot be completely devoid of
gas, the true extent along the l.o.s.\ should be significantly greater.

We have reported a similar subtle channel in the merging cluster A520
\citep{2016ApJ...833...99W}. There, it was aligned with an apparent direction
of a secondary subcluster merger. An intriguing possibility is that these
channels are examples of a plasma depletion layer (PDL) --- a feature observed
when the magnetic field gets stretched and amplified to values where its
energy density becomes comparable to thermal pressure of its host plasma. This
happens, for example, when the solar wind drapes around a planetary
magnetosphere, gets amplified and squeezes the plasma out from the narrow
layer around the obstacle \citep{2004SSRv..111..185O}. A flow of magnetized
plasma around a cluster cool core was simulated, e.g., by
\cite{2008ApJ...677..993D}, and a similar draping phenomenon was predicted.
While they used a uniform magnetic field in the gas flow, a tangled field,
more representative of clusters, produces a similar end result
\citep{2013ApJ...762...78Z}. While cold fronts are obvious locations for
PDL, sheets and filaments of significantly amplified field can emerge in other
locations with coherent gas flows. \cite{2011ApJ...743...16Z} presented MHD
simulations of a sloshing core and traced the evolution of the magnetic
fields. In their Figure 23, there is a particularly illuminating example of a
plasma depletion phenomenon. A filament of an amplified magnetic field aligned
with the cold front, but located at a distance from it, is in pressure
equilibrium with the surrounding gas, but because the pressure contribution
from the amplified magnetic field is significant (30\% of thermal pressure ---
compared to the usual $\sim$1\%), its thermal pressure is reduced by that
amount essentially by squeezing the gas from the filament. This would produce
an X-ray feature just like the channel we see aligned with the cold front in
A2142. Our channel is located well within the sloshing region delineated
by the NW cold front, and coherent gas flows are easily expected throughout
this region. A possibly similar feature, though seen as an enhancement rather
than a depression in X-ray brightness, was reported near the cold front in the
Virgo core \citep{2016MNRAS.455..846W}.

The existence of such layers of draped magnetic fields around cold fronts have
long been proposed to explain the suppressed thermal conduction and diffusion
across the front and the front stability (\citealt{2000MNRAS.317L..57E};
\citealt{2001ApJ...551..160V}; \citetalias{2007PhR...443....1M}). The KH
instabilities at the southern front (\autoref{sec:kh}) allowed us to evaluate
the {\em effective}\/ ICM viscosity. If the layer that we see in A2142 indeed
has an amplified and ordered field and wraps around the southern front, it is
the likely underlying physical mechanism that regulates the growth of those KH
instabilities and determines that effective viscosity.

\section{Summary}
\label{sec:summary}

A2142 provides a laboratory to study several interesting effects in the
intracluster plasma and in cluster cool cores. It exhibits four cold fronts
--- three in the core (two of which were the initial discovery of cold fronts
in \citetalias{2000ApJ...541..542M}) and one 1~Mpc from the center, indicating
long-lived sloshing set off by a strong disturbance from a merger. In this
work, we have studied the three inner fronts using a 200~ks \chandra\ dataset.
For the southern front, we estimate the velocity of the tangential gas flow
inside the front from an estimate of the centripetal acceleration and obtain
$v=550\pm 190$~km/s ($M=0.36\pm 0.12$ w.r.t.\ the sound speed in the gas on
the hotter side of the front). The southern front is clearly disrupted by
Kelvin-Helmholtz instability, exhibiting two eddies separated by 55~kpc with
an amplitude of 6--7~kpc. This is only the second reported example of the clearly
observed KH eddies in the plane of the sky (the other one is A3667; other
reports of the KHI were based on interpreting the structure in the front
brightness profiles as eddies in projection). We compare the observed eddies
with the numeric study of the growth of KHI in the context of cluster cold
fronts by \cite{2013MNRAS.436.1721R}, who included isotropic viscosity in
their simulations. The A2142 eddies match the simulations if the isotropic,
Spitzer-like viscosity is suppressed by a factor at least 5. The velocity of
the gas flow is the biggest uncertainty in this estimate, but the viscosity
has to be lower than Spitzer even if we assume a $M=1$\/
flow. From the numeric comparison of the effects of isotropic Spitzer
viscosity and anisotropic Braginskii viscosity in the presence of gas sloshing
and stretching of the magnetic fields \citep{2015ApJ...798...90Z}, such a
suppressed effective isotropic viscosity is consistent with full Braginskii
anisotropic viscosity. Our viscosity constraints are in line with several
recent results for other clusters based on the KHI at cold fronts
(\citealt{2013ApJ...764...60R}; \citealt{2017ApJ...834...74S};
\citealt{2017MNRAS.467.3662I}) as well as on the observed details of gas
stripping for an infalling galaxy \citep{2017ApJ...848...27K}.

A2142 has a cool, dense peak, whose specific entropy index ($K\approx49$~keV~cm$^2$)
makes it a rare ``warm core,'' an intermediate case between the cool cores
with sharply peaked, low-entropy cores and non-cool-core clusters with flat
cores. The peak is offset from the BCG by 30~kpc. Once the cool sloshing
structure (that includes this peak, the inner cold front and the southern cold
front) is approximately subtracted using wavelet decomposition, we see that
the larger-scale emission in the core is well-centered on the BCG, confirming
the lensing result \citep{2008PASJ...60..345O} that the BCG is at the center
of the cluster gravitational potential. This is the largest observed offset
between the cool peak and the center of the potential for any cluster that
still exhibits a well-defined peak. The extreme sloshing in A2142 should have
displaced the former cool core from the center of the potential, which
facilitated its disruption, as simulated in \cite{2010ApJ...717..908Z}. The
displaced peak expands, loses the stability provided by gravity, and becomes
more susceptible to sloshing-induced mixing with the hotter gas. The BCG does
not show a strong AGN (exhibiting only a very weak radio source) and there is
no evidence for X-ray cavities, suggesting that the displaced peak has starved
its nucleus of the accreting gas for a significant period.

Finally, we detect an intriguing ``channel'' in the X-ray brightness,
$>$100~kpc long, $\sim$15~kpc wide, with a $\sim$10\% dip in brightness, that
appears to be aligned with the southern cold front. It is similar to the
channel we observed in A520 \citep{2016ApJ...833...99W} (though that channel
is aligned with the axis of a secondary merger, not with a cold front). The
channel should be a sheet of low-density gas seen edge-on. While some
non-obvious 3D gas distributions cannot be excluded based on the X-ray image
of this merging cluster, we think that a plausible explanation of this
feature is a plasma depletion layer. In such a layer, the stretched and
amplified magnetic field in the sloshing core may reach a pressure
comparable with the thermal pressure of the gas, squeezing the gas from the
layer. Such phenomena are observed when the solar wind flows around an
obstacle, and also seen in simulations of sloshing cluster cores that
include magnetic fields \citep{2011ApJ...743...16Z}. Such channels may provide
an interesting additional tool to study the intracluster magnetic fields.

\acknowledgments

We thank the referee for useful criticism and detailed comments. QHSW was
supported by Chandra grants GO3-14144Z, AR5-16013X and GO8-19114.

\bibliography{cluster}

\begin{thebibliography}{}
\expandafter\ifx\csname natexlab\endcsname\relax\def\natexlab#1{#1}\fi
\providecommand{\url}[1]{\href{#1}{#1}}
\providecommand{\dodoi}[1]{doi:~\href{http://doi.org/#1}{\nolinkurl{#1}}}
\providecommand{\doeprint}[1]{\href{http://ascl.net/#1}{\nolinkurl{http://ascl.net/#1}}}
\providecommand{\doarXiv}[1]{\href{https://arxiv.org/abs/#1}{\nolinkurl{https://arxiv.org/abs/#1}}}

\bibitem[{{Anders} \& {Grevesse}(1989)}]{1989GeCoA..53..197A}
{Anders}, E., \& {Grevesse}, N. 1989, \gca, 53, 197,
  \dodoi{10.1016/0016-7037(89)90286-X}

\bibitem[{{Ascasibar} \& {Markevitch}(2006)}]{2006ApJ...650..102A}
{Ascasibar}, Y., \& {Markevitch}, M. 2006, \apj, 650, 102,
  \dodoi{10.1086/506508}

\bibitem[{{Buote} {et~al.}(2005){Buote}, {Humphrey}, \&
  {Stocke}}]{2005ApJ...630..750B}
{Buote}, D.~A., {Humphrey}, P.~J., \& {Stocke}, J.~T. 2005, \apj, 630, 750,
  \dodoi{10.1086/432045}

\bibitem[{{Cavagnolo} {et~al.}(2009){Cavagnolo}, {Donahue}, {Voit}, \&
  {Sun}}]{2009ApJS..182...12C}
{Cavagnolo}, K.~W., {Donahue}, M., {Voit}, G.~M., \& {Sun}, M. 2009, \apjs,
  182, 12, \dodoi{10.1088/0067-0049/182/1/12}

\bibitem[{{Churazov} {et~al.}(2003){Churazov}, {Forman}, {Jones}, \&
  {B{\"o}hringer}}]{2003ApJ...590..225C}
{Churazov}, E., {Forman}, W., {Jones}, C., \& {B{\"o}hringer}, H. 2003, \apj,
  590, 225, \dodoi{10.1086/374923}

\bibitem[{{Clarke} {et~al.}(2004){Clarke}, {Blanton}, \&
  {Sarazin}}]{2004ApJ...616..178C}
{Clarke}, T.~E., {Blanton}, E.~L., \& {Sarazin}, C.~L. 2004, \apj, 616, 178,
  \dodoi{10.1086/424911}

\bibitem[{{Dupke} {et~al.}(2007){Dupke}, {White}, \&
  {Bregman}}]{2007ApJ...671..181D}
{Dupke}, R., {White}, III, R.~E., \& {Bregman}, J.~N. 2007, \apj, 671, 181,
  \dodoi{10.1086/522194}

\bibitem[{{Dursi} \& {Pfrommer}(2008)}]{2008ApJ...677..993D}
{Dursi}, L.~J., \& {Pfrommer}, C. 2008, \apj, 677, 993, \dodoi{10.1086/529371}

\bibitem[{{Ettori} \& {Fabian}(2000)}]{2000MNRAS.317L..57E}
{Ettori}, S., \& {Fabian}, A.~C. 2000, \mnras, 317, L57,
  \dodoi{10.1046/j.1365-8711.2000.03899.x}

\bibitem[{{Ettori} {et~al.}(2013){Ettori}, {Gastaldello}, {Gitti},
  {O'Sullivan}, {Gaspari}, {Brighenti}, {David}, \&
  {Edge}}]{2013A&A...555A..93E}
{Ettori}, S., {Gastaldello}, F., {Gitti}, M., {et~al.} 2013, \aap, 555, A93,
  \dodoi{10.1051/0004-6361/201321107}

\bibitem[{{Ghizzardi} {et~al.}(2010){Ghizzardi}, {Rossetti}, \&
  {Molendi}}]{2010A&A...516A..32G}
{Ghizzardi}, S., {Rossetti}, M., \& {Molendi}, S. 2010, \aap, 516, A32,
  \dodoi{10.1051/0004-6361/200912496}

\bibitem[{{Giacintucci} {et~al.}(2017){Giacintucci}, {Markevitch}, {Cassano},
  {Venturi}, {Clarke}, \& {Brunetti}}]{2017ApJ...841...71G}
{Giacintucci}, S., {Markevitch}, M., {Cassano}, R., {et~al.} 2017, \apj, 841,
  71, \dodoi{10.3847/1538-4357/aa7069}

\bibitem[{{Hamer} {et~al.}(2012){Hamer}, {Edge}, {Swinbank}, {Wilman},
  {Russell}, {Fabian}, {Sanders}, \& {Salom{\'e}}}]{2012MNRAS.421.3409H}
{Hamer}, S.~L., {Edge}, A.~C., {Swinbank}, A.~M., {et~al.} 2012, \mnras, 421,
  3409, \dodoi{10.1111/j.1365-2966.2012.20566.x}

\bibitem[{{Hickox} \& {Markevitch}(2006)}]{2006ApJ...645...95H}
{Hickox}, R.~C., \& {Markevitch}, M. 2006, \apj, 645, 95,
  \dodoi{10.1086/504070}

\bibitem[{{Ichinohe} {et~al.}(2017){Ichinohe}, {Simionescu}, {Werner}, \&
  {Takahashi}}]{2017MNRAS.467.3662I}
{Ichinohe}, Y., {Simionescu}, A., {Werner}, N., \& {Takahashi}, T. 2017,
  \mnras, 467, 3662, \dodoi{10.1093/mnras/stx280}

\bibitem[{{Johnson}(2011)}]{2011PhDT........14J}
{Johnson}, R. 2011, PhD thesis, Dartmouth College

\bibitem[{{Kalberla} {et~al.}(2005){Kalberla}, {Burton}, {Hartmann}, {Arnal},
  {Bajaja}, {Morras}, \& {P{\"o}ppel}}]{2005A&A...440..775K}
{Kalberla}, P.~M.~W., {Burton}, W.~B., {Hartmann}, D., {et~al.} 2005, \aap,
  440, 775, \dodoi{10.1051/0004-6361:20041864}

\bibitem[{{Keshet} {et~al.}(2010){Keshet}, {Markevitch}, {Birnboim}, \&
  {Loeb}}]{2010ApJ...719L..74K}
{Keshet}, U., {Markevitch}, M., {Birnboim}, Y., \& {Loeb}, A. 2010, \apjl, 719,
  L74, \dodoi{10.1088/2041-8205/719/1/L74}

\bibitem[{{Kraft} {et~al.}(2017){Kraft}, {Roediger}, {Machacek}, {Forman},
  {Nulsen}, {Jones}, {Churazov}, {Randall}, {Su}, \&
  {Sheardown}}]{2017ApJ...848...27K}
{Kraft}, R.~P., {Roediger}, E., {Machacek}, M., {et~al.} 2017, \apj, 848, 27,
  \dodoi{10.3847/1538-4357/aa8a6e}

\bibitem[{{Lyutikov}(2006)}]{2006MNRAS.373...73L}
{Lyutikov}, M. 2006, \mnras, 373, 73, \dodoi{10.1111/j.1365-2966.2006.10835.x}

\bibitem[{{Machacek} {et~al.}(2005){Machacek}, {Dosaj}, {Forman}, {Jones},
  {Markevitch}, {Vikhlinin}, {Warmflash}, \& {Kraft}}]{2005ApJ...621..663M}
{Machacek}, M., {Dosaj}, A., {Forman}, W., {et~al.} 2005, \apj, 621, 663,
  \dodoi{10.1086/427548}

\bibitem[{{Markevitch} {et~al.}(2002){Markevitch}, {Gonzalez}, {David},
  {Vikhlinin}, {Murray}, {Forman}, {Jones}, \& {Tucker}}]{2002ApJ...567L..27M}
{Markevitch}, M., {Gonzalez}, A.~H., {David}, L., {et~al.} 2002, \apjl, 567,
  L27, \dodoi{10.1086/339619}

\bibitem[{{Markevitch} \& {Vikhlinin}(2007)}]{2007PhR...443....1M}
{Markevitch}, M., \& {Vikhlinin}, A. 2007, \physrep, 443, 1,
  \dodoi{10.1016/j.physrep.2007.01.001}

\bibitem[{{Markevitch} {et~al.}(2001){Markevitch}, {Vikhlinin}, \&
  {Mazzotta}}]{2001ApJ...562L.153M}
{Markevitch}, M., {Vikhlinin}, A., \& {Mazzotta}, P. 2001, \apjl, 562, L153,
  \dodoi{10.1086/337973}

\bibitem[{{Markevitch} {et~al.}(2000){Markevitch}, {Ponman}, {Nulsen}, {Bautz},
  {Burke}, {David}, {Davis}, {Donnelly}, {Forman}, {Jones}, {Kaastra},
  {Kellogg}, {Kim}, {Kolodziejczak}, {Mazzotta}, {Pagliaro}, {Patel}, {Van
  Speybroeck}, {Vikhlinin}, {Vrtilek}, {Wise}, \& {Zhao}}]{2000ApJ...541..542M}
{Markevitch}, M., {Ponman}, T.~J., {Nulsen}, P.~E.~J., {et~al.} 2000, \apj,
  541, 542, \dodoi{10.1086/309470}

\bibitem[{{Markevitch} {et~al.}(2003){Markevitch}, {Bautz}, {Biller}, {Butt},
  {Edgar}, {Gaetz}, {Garmire}, {Grant}, {Green}, {Juda}, {Plucinsky},
  {Schwartz}, {Smith}, {Vikhlinin}, {Virani}, {Wargelin}, \&
  {Wolk}}]{2003ApJ...583...70M}
{Markevitch}, M., {Bautz}, M.~W., {Biller}, B., {et~al.} 2003, \apj, 583, 70,
  \dodoi{10.1086/345347}

\bibitem[{{Mazzotta} {et~al.}(2002){Mazzotta}, {Fusco-Femiano}, \&
  {Vikhlinin}}]{2002ApJ...569L..31M}
{Mazzotta}, P., {Fusco-Femiano}, R., \& {Vikhlinin}, A. 2002, \apjl, 569, L31,
  \dodoi{10.1086/340481}

\bibitem[{{Mazzotta} {et~al.}(2001){Mazzotta}, {Markevitch}, {Vikhlinin},
  {Forman}, {David}, \& {van Speybroeck}}]{2001ApJ...555..205M}
{Mazzotta}, P., {Markevitch}, M., {Vikhlinin}, A., {et~al.} 2001, \apj, 555,
  205, \dodoi{10.1086/321484}

\bibitem[{{Million} {et~al.}(2010){Million}, {Allen}, {Werner}, \&
  {Taylor}}]{2010MNRAS.405.1624M}
{Million}, E.~T., {Allen}, S.~W., {Werner}, N., \& {Taylor}, G.~B. 2010,
  \mnras, 405, 1624, \dodoi{10.1111/j.1365-2966.2010.16596.x}

\bibitem[{{Oegerle} {et~al.}(1995){Oegerle}, {Hill}, \&
  {Fitchett}}]{1995AJ....110...32O}
{Oegerle}, W.~R., {Hill}, J.~M., \& {Fitchett}, M.~J. 1995, \aj, 110, 32,
  \dodoi{10.1086/117495}

\bibitem[{{{\O}ieroset} {et~al.}(2004){{\O}ieroset}, {Mitchell}, {Phan}, {Lin},
  {Crider}, \& {Acu{\~n}a}}]{2004SSRv..111..185O}
{{\O}ieroset}, M., {Mitchell}, D.~L., {Phan}, T.~D., {et~al.} 2004, \ssr, 111,
  185, \dodoi{10.1023/B:SPAC.0000032715.69695.9c}

\bibitem[{{Okabe} \& {Umetsu}(2008)}]{2008PASJ...60..345O}
{Okabe}, N., \& {Umetsu}, K. 2008, \pasj, 60, 345,
  \dodoi{10.1093/pasj/60.2.345}

\bibitem[{{Owers} {et~al.}(2009){Owers}, {Nulsen}, {Couch}, \&
  {Markevitch}}]{2009ApJ...704.1349O}
{Owers}, M.~S., {Nulsen}, P.~E.~J., {Couch}, W.~J., \& {Markevitch}, M. 2009,
  \apj, 704, 1349, \dodoi{10.1088/0004-637X/704/2/1349}

\bibitem[{{Roediger} {et~al.}(2011){Roediger}, {Br{\"u}ggen}, {Simionescu},
  {B{\"o}hringer}, {Churazov}, \& {Forman}}]{2011MNRAS.413.2057R}
{Roediger}, E., {Br{\"u}ggen}, M., {Simionescu}, A., {et~al.} 2011, \mnras,
  413, 2057, \dodoi{10.1111/j.1365-2966.2011.18279.x}

\bibitem[{{Roediger} {et~al.}(2013{\natexlab{a}}){Roediger}, {Kraft}, {Forman},
  {Nulsen}, \& {Churazov}}]{2013ApJ...764...60R}
{Roediger}, E., {Kraft}, R.~P., {Forman}, W.~R., {Nulsen}, P.~E.~J., \&
  {Churazov}, E. 2013{\natexlab{a}}, \apj, 764, 60,
  \dodoi{10.1088/0004-637X/764/1/60}

\bibitem[{{Roediger} {et~al.}(2013{\natexlab{b}}){Roediger}, {Kraft}, {Nulsen},
  {Churazov}, {Forman}, {Br{\"u}ggen}, \& {Kokotanekova}}]{2013MNRAS.436.1721R}
{Roediger}, E., {Kraft}, R.~P., {Nulsen}, P., {et~al.} 2013{\natexlab{b}},
  \mnras, 436, 1721, \dodoi{10.1093/mnras/stt1691}

\bibitem[{{Roediger} {et~al.}(2012){Roediger}, {Lovisari}, {Dupke},
  {Ghizzardi}, {Br{\"u}ggen}, {Kraft}, \& {Machacek}}]{2012MNRAS.420.3632R}
{Roediger}, E., {Lovisari}, L., {Dupke}, R., {et~al.} 2012, \mnras, 420, 3632,
  \dodoi{10.1111/j.1365-2966.2011.20287.x}

\bibitem[{{Rossetti} {et~al.}(2013){Rossetti}, {Eckert}, {De Grandi},
  {Gastaldello}, {Ghizzardi}, {Roediger}, \& {Molendi}}]{2013A&A...556A..44R}
{Rossetti}, M., {Eckert}, D., {De Grandi}, S., {et~al.} 2013, \aap, 556, A44,
  \dodoi{10.1051/0004-6361/201321319}

\bibitem[{{Sarazin}(1988)}]{1988xrec.book.....S}
{Sarazin}, C.~L. 1988, {X-ray emission from clusters of galaxies}

\bibitem[{{Simionescu} {et~al.}(2010){Simionescu}, {Werner}, {Forman},
  {Miller}, {Takei}, {B{\"o}hringer}, {Churazov}, \&
  {Nulsen}}]{2010MNRAS.405...91S}
{Simionescu}, A., {Werner}, N., {Forman}, W.~R., {et~al.} 2010, \mnras, 405,
  91, \dodoi{10.1111/j.1365-2966.2010.16450.x}

\bibitem[{{Simionescu} {et~al.}(2012){Simionescu}, {Werner}, {Urban}, {Allen},
  {Fabian}, {Sanders}, {Mantz}, {Nulsen}, \& {Takei}}]{2012ApJ...757..182S}
{Simionescu}, A., {Werner}, N., {Urban}, O., {et~al.} 2012, \apj, 757, 182,
  \dodoi{10.1088/0004-637X/757/2/182}

\bibitem[{{Spitzer}(1962)}]{1962pfig.book.....S}
{Spitzer}, L. 1962, {Physics of Fully Ionized Gases}

\bibitem[{{Su} {et~al.}(2017){Su}, {Kraft}, {Roediger}, {Nulsen}, {Forman},
  {Churazov}, {Randall}, {Jones}, \& {Machacek}}]{2017ApJ...834...74S}
{Su}, Y., {Kraft}, R.~P., {Roediger}, E., {et~al.} 2017, \apj, 834, 74,
  \dodoi{10.3847/1538-4357/834/1/74}

\bibitem[{{Tittley} \& {Henriksen}(2005)}]{2005ApJ...618..227T}
{Tittley}, E.~R., \& {Henriksen}, M. 2005, \apj, 618, 227,
  \dodoi{10.1086/425952}

\bibitem[{{Venturi} {et~al.}(2017){Venturi}, {Rossetti}, {Brunetti},
  {Farnsworth}, {Gastaldello}, {Giacintucci}, {Lal}, {Rudnick}, {Shimwell},
  {Eckert}, {Molendi}, \& {Owers}}]{2017A&A...603A.125V}
{Venturi}, T., {Rossetti}, M., {Brunetti}, G., {et~al.} 2017, \aap, 603, A125,
  \dodoi{10.1051/0004-6361/201630014}

\bibitem[{{Vikhlinin}(2011)}]{2011scgg.conf...53V}
{Vikhlinin}, A. 2011, in Structure in Clusters and Groups of Galaxies in the
  Chandra Era, ed. J.~{Vrtilek} \& P.~J. {Green}, 53

\bibitem[{{Vikhlinin} {et~al.}(1994){Vikhlinin}, {Forman}, \&
  {Jones}}]{1994ApJ...435..162V}
{Vikhlinin}, A., {Forman}, W., \& {Jones}, C. 1994, \apj, 435, 162,
  \dodoi{10.1086/174802}

\bibitem[{{Vikhlinin} {et~al.}(2001{\natexlab{a}}){Vikhlinin}, {Markevitch}, \&
  {Murray}}]{2001ApJ...551..160V}
{Vikhlinin}, A., {Markevitch}, M., \& {Murray}, S.~S. 2001{\natexlab{a}}, \apj,
  551, 160, \dodoi{10.1086/320078}

\bibitem[{{Vikhlinin} {et~al.}(2001{\natexlab{b}}){Vikhlinin}, {Markevitch}, \&
  {Murray}}]{2001ApJ...549L..47V}
---. 2001{\natexlab{b}}, \apjl, 549, L47, \dodoi{10.1086/319126}

\bibitem[{{Vikhlinin} {et~al.}(2005){Vikhlinin}, {Markevitch}, {Murray},
  {Jones}, {Forman}, \& {Van Speybroeck}}]{2005ApJ...628..655V}
{Vikhlinin}, A., {Markevitch}, M., {Murray}, S.~S., {et~al.} 2005, \apj, 628,
  655, \dodoi{10.1086/431142}

\bibitem[{{Vikhlinin} {et~al.}(1998){Vikhlinin}, {McNamara}, {Forman}, {Jones},
  {Quintana}, \& {Hornstrup}}]{1998ApJ...502..558V}
{Vikhlinin}, A., {McNamara}, B.~R., {Forman}, W., {et~al.} 1998, \apj, 502,
  558, \dodoi{10.1086/305951}

\bibitem[{{Walker} {et~al.}(2017){Walker}, {Hlavacek-Larrondo},
  {Gendron-Marsolais}, {Fabian}, {Intema}, {Sanders}, {Bamford}, \& {van
  Weeren}}]{2017MNRAS.468.2506W}
{Walker}, S.~A., {Hlavacek-Larrondo}, J., {Gendron-Marsolais}, M., {et~al.}
  2017, \mnras, 468, 2506, \dodoi{10.1093/mnras/stx640}

\bibitem[{{Walker} {et~al.}(2016){Walker}, {Sanders}, \&
  {Fabian}}]{2016MNRAS.461..684W}
{Walker}, S.~A., {Sanders}, J.~S., \& {Fabian}, A.~C. 2016, \mnras, 461, 684,
  \dodoi{10.1093/mnras/stw1367}

\bibitem[{{Wang} {et~al.}(2016){Wang}, {Markevitch}, \&
  {Giacintucci}}]{2016ApJ...833...99W}
{Wang}, Q.~H.~S., {Markevitch}, M., \& {Giacintucci}, S. 2016, \apj, 833, 99,
  \dodoi{10.3847/1538-4357/833/1/99}

\bibitem[{{Werner} {et~al.}(2016{\natexlab{a}}){Werner}, {Zhuravleva},
  {Canning}, {Allen}, {King}, {Sanders}, {Simionescu}, {Taylor}, {Morris}, \&
  {Fabian}}]{2016MNRAS.460.2752W}
{Werner}, N., {Zhuravleva}, I., {Canning}, R.~E.~A., {et~al.}
  2016{\natexlab{a}}, \mnras, 460, 2752, \dodoi{10.1093/mnras/stw1171}

\bibitem[{{Werner} {et~al.}(2016{\natexlab{b}}){Werner}, {ZuHone},
  {Zhuravleva}, {Ichinohe}, {Simionescu}, {Allen}, {Markevitch}, {Fabian},
  {Keshet}, {Roediger}, {Ruszkowski}, \& {Sanders}}]{2016MNRAS.455..846W}
{Werner}, N., {ZuHone}, J.~A., {Zhuravleva}, I., {et~al.} 2016{\natexlab{b}},
  \mnras, 455, 846, \dodoi{10.1093/mnras/stv2358}

\bibitem[{{ZuHone} {et~al.}(2015){ZuHone}, {Kunz}, {Markevitch}, {Stone}, \&
  {Biffi}}]{2015ApJ...798...90Z}
{ZuHone}, J.~A., {Kunz}, M.~W., {Markevitch}, M., {Stone}, J.~M., \& {Biffi},
  V. 2015, \apj, 798, 90, \dodoi{10.1088/0004-637X/798/2/90}

\bibitem[{{ZuHone} {et~al.}(2013){ZuHone}, {Markevitch}, {Brunetti}, \&
  {Giacintucci}}]{2013ApJ...762...78Z}
{ZuHone}, J.~A., {Markevitch}, M., {Brunetti}, G., \& {Giacintucci}, S. 2013,
  \apj, 762, 78, \dodoi{10.1088/0004-637X/762/2/78}

\bibitem[{{ZuHone} {et~al.}(2010){ZuHone}, {Markevitch}, \&
  {Johnson}}]{2010ApJ...717..908Z}
{ZuHone}, J.~A., {Markevitch}, M., \& {Johnson}, R.~E. 2010, \apj, 717, 908,
  \dodoi{10.1088/0004-637X/717/2/908}

\bibitem[{{ZuHone} {et~al.}(2011){ZuHone}, {Markevitch}, \&
  {Lee}}]{2011ApJ...743...16Z}
{ZuHone}, J.~A., {Markevitch}, M., \& {Lee}, D. 2011, \apj, 743, 16,
  \dodoi{10.1088/0004-637X/743/1/16}

\end{thebibliography}

\end{document}